\documentclass[aps,prd,amsfonts,amsmath,preprint,nofootinbib]{revtex4-2}

\usepackage[utf8]{inputenc} 
\usepackage{graphicx,xcolor,overpic,mathtools}
\usepackage[colorlinks]{hyperref}
\hypersetup{ 
	colorlinks=true, 
	linkcolor=black, 
	filecolor=black, 
	citecolor = blue,       
	urlcolor=blue, 
} 
\newcommand{\beq}{\begin{equation}}
\newcommand{\eeq}{\end{equation}}

\usepackage{upgreek}
\usepackage{csquotes}
\numberwithin{equation}{section}

\usepackage[title]{appendix}
\usepackage{verbatim}

\def\beq{\begin{equation}} 
\def\eeq{\end{equation}}
\def\beqa{\begin{eqnarray}}
\def\eeqa{\end{eqnarray}}

\usepackage{amsthm,amsmath,amssymb}

\PassOptionsToPackage{normalem}{ulem}
\usepackage{ulem} 
\usepackage{sidenotes}
\usepackage{bm}
\usepackage{url}
\usepackage{physics}
\usepackage{xfrac}
\usepackage[makeroom]{cancel}
\usepackage{tensor}
\usepackage{mathrsfs}
\usepackage{slashed}
\usepackage{tikz}
\usepackage[mode=buildnew]{standalone}
\usepackage{bbm}
\usepackage{cleveref}
\crefname{section}{Sec.}{sec.}
\crefname{figure}{Fig.}{Fig.}
\Crefname{section}{Section}{Sections}
\usepackage{caption,subcaption}
\newcommand{\lag}{\mathcal{L}}

\NewDocumentCommand{\evat}{sO{\bigg}mm}{%
  \IfBooleanTF{#1}
   {\mleft. #3 \mright|_{#4}}
   {#3#2|_{#4}}%
}

\begin{document}
\count\footins = 1000 %avoids footnotes over bottom page numbers

\title{Effective Actions for Domain Wall Dynamics }

\author{Jose J. Blanco-Pillado$^{1,2,3}$}
\author{Alberto Garc\'ia Mart\'in-Caro$^{2,3}$}
\author{Daniel Jim\'enez-Aguilar$^4$}
\author{Jose M. Queiruga$^{5,6}$}

\affiliation{$^1$ IKERBASQUE, Basque Foundation for Science, 48011, Bilbao, Spain,}
\affiliation{$^2$ EHU Quantum Center, University of the Basque Country, UPV/EHU,}
\affiliation{$^3$ Department of Physics, University of the Basque Country, UPV/EHU, 48080, Bilbao, Spain,}
 \affiliation{$^4$ Institute of Cosmology, Department of Physics and Astronomy, Tufts University, Medford, MA 02155, USA,}
\affiliation{$^5$ Department of Applied Mathematics, University of Salamanca, Casas del Parque 2}
\affiliation{$^6$ Institute of Fundamental Physics and Mathematics, University of Salamanca,
Plaza de la Merced 1, 37008 - Salamanca, Spain}

%\vspace{-5 cm}

\begin{abstract}
We introduce a systematic method to derive the effective action for domain walls directly from the scalar field theory that gives rise to their solitonic solutions. The effective action for the Goldstone mode, which characterizes the soliton’s position, is shown to consist of the Nambu-Goto action supplemented by higher-order curvature invariants associated to its worldvolume metric. Our approach constrains the corrections to a finite set of Galileon terms, specifying both their functional forms and the procedure to compute their coefficients. We do a collection of tests across various models in $2+1$ and $3+1$ dimensions that confirm the validity of this framework. Additionally, the method is extended to include bound scalar fields living on the worldsheet, along with their couplings to the Goldstone mode. These interactions reveal a universal non-minimal coupling of these scalar fields to the Ricci scalar on the worldsheet. A significant consequence of this coupling is the emergence of a parametric instability, driven by interactions between the bound states and the Goldstone mode.    
\end{abstract}

\maketitle

\newpage

\section{Introduction}

Soliton configurations are of fundamental significance in various areas of physics, including, among others, 
condensed matter, particle physics and cosmology (see, for example, \cite{BISHOP19801,rajaraman1982solitons,Vilenkin:2000jqa,Manton:2004tk}). 
Understanding their behaviour and interactions is an essential aspect of research in these fields as their dynamics 
largely determine the majority of their observational signatures.

In this context, solitons typically emerge as classical solutions of a non-linear field theory with complicated equations of 
motion. While solving the complete set of equations for all fields in these theories is possible, it often 
requires substantial computational resources. In many cases, however, the primary interest lies in understanding 
the soliton's behavior as a collective entity, rather than the dynamics of each individual field.

One effective approach to this problem is to reduce the number of degrees of freedom by identifying those 
that play a significant role in the phenomenon of interest. This idea has inspired several different attempts 
to capture the relevant physics of these problems. A successful strategy for doing so is the use of effective 
field theories, which have been widely applied across various areas of physics. This approach involves writing 
down the most general action for the relevant degrees of freedom, constrained by the symmetries of the 
system. The challenge in this method is to calculate the coefficients of all the terms allowed by this 
procedure\footnote{An alternative approach would be to capture the relevant behavior by employing a 
suitable ansatz for the soliton field that parametrizes the specific type of dynamics of interest. This 
method is known as the collective coordinate method or moduli space approximation, and it has been 
successfully applied numerous times in soliton physics (See for example \cite{Gervais:1974dc,Manton:1977er,Rice83}).}.

In principle, one would like to understand whether this effective action can be derived explicitly from the 
original field theory by integrating out the irrelevant degrees of freedom. This process not only generates new 
interaction terms for the low-energy degrees of freedom but also provides, in principle, a means to compute 
the corresponding coefficients.

In this paper we aim to apply this procedure to the dynamics of extended solitons. To facilitate
this analysis, it is essential to first identify the types of excitations encountered around solitons. A
useful approach to classify these excitations is based on their physical origin.

The first type comprises universal excitations that describe the arbitrary position of the center of the soliton
in spacetime. These excitations are referred to as Nambu-Goldstone modes, as they correspond to the massless
mode due to the breaking of translational invariance associated with the soliton solution. Consequently, 
these modes can be interpreted as massless excitations of the rigid soliton.

The second category is more model-dependent and may or may not be present in specific models. These excitations 
are typically associated with the internal structure of the soliton. From the soliton's perspective, these 
excitations are massive and are generally localized within a region of the order of the soliton's thickness.

Finally, we should also consider additional modes that are not bound to the soliton, representing 
the degrees of freedom propagating in the vacuum away from the soliton. These modes also interact 
with the soliton and may lead to significant effects, as they can be radiated away from the soliton. 
The energy of these modes are larger compared to the bound states described above. It is precisely 
due to this separation of scales that one can hope to use the Effective Action techniques
to obtain an action that encapsulates the dynamics of the low energy degrees of freedom. 

For extended solitons, those like domain walls in $2+1$ dimensions, strings, or domain walls in $3+1$ dimensions, the 
lowest energy modes are the ones that describe the motion of the objects in spacetime. Determining the correct effective 
action for soliton motion is crucial for extracting reliable information about their evolution in any context. In this 
paper, our primary interest is the study of these motions in connection to the existence of topological defects in 
cosmology. Numerous high energy extensions beyond the Standard Model predict the formation of topological defects 
in the Early Universe \cite{Kibble:1976sj,Vilenkin:2000jqa}. While some of these models have been explored 
through cosmological field theory simulations, these numerical implementations are computationally 
expensive. Thus, it is desirable 
to derive an effective action for these objects that accurately and efficiently identifies their evolution 
in field theory. This will allow for larger or longer simulations which presumably would lead to more 
accurate predictions of the model in a cosmological context.

It was shown some time ago \cite{Forster:1974ga,1980ZPhyB..36..329D} that the effective action for the zero modes 
describing soliton displacements is, at leading order, identical to the Nambu-Goto (NG) action \cite{nambu1970lectures,Goto:1971ce}. Subsequent 
analyses of higher-order corrections to this action have generated some debate regarding the appropriate terms to include 
\cite{Maeda:1987pd, Gregory:1988qv,Gregory:1989gg,Gregory:1990pm,Anderson:1997ip}. In particular, several previous studies 
have specifically examined the domain wall case, offering different insights to the problem we focus on here. However, the 
final results are not always agreeing with each other \cite{Letelier:1990jt,Gregory:1989gg,Silveira:1993am,Larsen:1993np,Carter:1994ag}. 
In this work, we directly derive these higher-order corrections from the field theory model that supports soliton 
solutions. Moreover, the method we employ to obtain these corrections naturally incorporates the coupling of the zero modes to massive modes.

The significance of bound states in soliton dynamics and their potential relevance to realistic scenarios has only recently begun to attract some attention in the soliton community \cite{Blanco-Pillado:2020smt,Blanco-Pillado:2021jad,Alonso-Izquierdo:2024tjc,Adam:2019xuc,Alonso-Izquierdo:2024fpw,Krusch:2024vuy}. Some recent claims suggest that these bound states could impact the dynamics of cosmic strings \cite{Hindmarsh:2021mnl}. These ideas have motivated more detailed investigations that compared the evolutions predicted within the Nambu-Goto action and the results from field theory \cite{Blanco-Pillado:2023sap}. Although the role of these modes in the cosmological evolution of strings remains uncertain, the study of soliton excitations is valuable in its own right and could find applications in other areas of physics. 

In this paper, we consider a simpler system: the dynamics of domain walls in a scalar field theory in $2+1$ and $3+1$ dimensions. One significant advantage of these models is their simplicity. Domain walls can be easily simulated and evolved in lattice simulations with moderate computational resources. Additionally, many of these models offer analytic solutions for both solitons and their excitations, providing better control over the identification of excitations in numerical simulations. As we demonstrate in this paper, these models provide an ideal testing ground for computing the effective action and its coupling to various modes allowing for a precise comparison of the evolution in field theory with the effective action.

Moreover, these models offer insights into some recent early universe scenarios. While 
domain walls were traditionally viewed as problematic for cosmology due to their potential to dominate 
the universe's energy density 
and yield predictions inconsistent with observations \cite{Zeldovich:1974uw}, it was realized early on that these 
models could remain viable if mechanisms existed to cause the walls' disappearance. Interestingly, recent work 
suggests that the eventual demise of domain walls could make these models more interesting, as this process could 
leave observable signatures \cite{Ferreira:2022zzo,Dankovsky:2024zvs}. Specifically, the collapse of domain walls 
could lead to gravitational wave generation and primordial black hole formation \cite{Gouttenoire:2023ftk,Ferreira:2024eru,Dunsky:2024zdo}. Although some preliminary studies of these scenarios have been conducted, the lack of a detailed understanding of soliton effective actions introduces uncertainties. In this paper, we will examine an aspect of domain wall evolution that could be important during their final stages of collapse.\\

The rest of the paper is organized as follows. In \cref{Sec:models}, we briefly introduce the field theory models under
investigation and describe the type of low energy excitations that they have.
In \cref{Sec:general_discussion}, we discuss the general method for constructing an effective action and how to apply 
it to the problem of domain wall dynamics. In \cref{Sec:deriving_action}, we explain the method we follow in this paper 
to compute the effective action for the domain walls from the underlying field theory. 
\cref{sec:numerics_effective} focuses on the precise dynamics of the collapse of circular and spherical domain walls 
and compares the results from our effective action with those from direct integration of the 
field theory equations in a lattice.  This section also explores the dependence of the 
effective action and the evolution of the walls on the underlying field theory model. In \cref{Sec:profile_corrections} we examine
the relationship between curvature corrections and modifications to the soliton profile. In \cref{sec:boundstate_eff}, 
we extend the effective action to include couplings to bound states. In \cref{Sec:numerical_boundstate}, we study two key consequences of the bound state's coupling to the Goldstone mode: the emergence of a parametric instability and the resulting backreaction on the dynamics of a collapsing wall. We conclude in \cref{Sec:Conclusions} with a summary and  potential future research directions. Finally, detailed derivations supporting the conclusions in the main text are provided in the appendices for clarity and completeness.

%%%%%%%%%%%%%%%%%%%%%%%%%%%%%%%%%%%%%%%%%%%
\section{The domain wall in field theory}
\label{Sec:models}
%%%%%%%%%%%%%%%%%%%%%%%%%%%%%%%%%%%%%%%%%%%

In this section, we outline the field theory models under consideration and explain 
how domain wall solitons emerge as solutions within these models. To provide a 
concrete example, we focus on the canonical model of the simplest scalar field 
soliton based on the $\lambda \phi^4$ scalar field theory. While most of the 
subsequent discussion remains independent of the specific model, as we 
will note in the following section, the exact form of the coefficients in the 
low-energy effective action that we find is model-dependent.

Let us then consider the following scalar field theory,
\beq
\label{FT-action}
S = \int{ d^{D+1}x \left[-\frac{1}{2} \partial_{\mu} \phi \partial^{\mu} \phi - \frac{\lambda}{4} \left(\phi^2 - \eta^2\right)^2\right]}~,
\eeq
where the number of spatial dimensions $D$ has been left unspecified since
we will discuss several different cases, namely in $2$ and $3$ spatial dimensions\footnote{In this
work we will use the mostly positive signature and a flat spacetime metric, $\eta_{\mu \nu}$.}.

This theory is known to support soliton solutions that interpolate between the different vacua of the
potential. The static solution for these solitons can be expressed as
\beq
\label{kink-f4}
\phi_K(x) = \eta \tanh \left[l^{-1} (z - \psi)\right]\,,
\eeq
where $\psi$ represents the position of the soliton along the $z$ coordinate, and 
$l=\sqrt{\frac{2}{\lambda} }\left(\frac{1}{\eta}\right)$ is the 
characteristic width of the soliton profile. In $1+1$ dimensions, this solution is typically 
referred to as a kink. It localizes most of its energy in a small region around $\psi$ and behaves 
like a point particle of mass $M_K=\frac{2\sqrt{2}}{3} \sqrt{\lambda} \eta^3$. However, this type of solution can also exist in higher-dimensional 
theories, giving rise to extended soliton configurations. For instance, in one additional dimension, 
domain walls manifest as string-like structures. These {\it domain wall strings} possess an energy 
density per unit length —essentially their tension— equal to the mass of the kink, quantified as before\footnote{Note however that the units of the quantities given here depend on the dimensionality of the theory in which they appear.}, $\mu = \frac{2\sqrt{2}}{3} \sqrt{\lambda} \eta^3$. These extended 
solitons and their dynamics form the primary focus of this study.

As discussed in the introduction, the spectrum of perturbations around this type of soliton 
solution can be obtained analytically in simple cases, allowing for a detailed characterization 
of their dynamics and interactions (see, for example, \cite{Vachaspati:2006zz,Blanco-Pillado:2022rad}). In particular, the analytic expressions for the mode wavefunctions offer valuable insights into their behavior.

For our purposes, it is sufficient to focus on the most relevant excitation modes. The spectrum 
begins with a massless mode that corresponds to the rigid translation of the soliton, evident 
from the fact that the soliton solution is parameterized by an arbitrary position, $\psi$. In higher 
dimensions, these massless excitations manifest as propagating {\it wiggles} along the soliton longitudinal
dimensions.

The second type of excitation encountered in these models is a bound state, a massive mode localized 
near the soliton's core. In the case of the kink, this mode is referred to as the {\it shape mode}, 
representing a deformation of the soliton profile. The existence of such modes is model-dependent, 
but in the $\lambda \phi^4$ theory, a shape mode exists with a mass of $m_s = \frac{\sqrt{3}}{2} m$, where $m^2= 2\lambda \eta^2$ is the mass of the propagating modes in the vacuum of the theory.

Finally, the model also supports scattering states, which are excitations that interact with the 
soliton and enable energy dissipation into the vacuum through radiation or other decay processes. We will
only briefly touch upon these last modes in this paper where we focus in the dynamical effects
of the other two types of modes.

%%%%%%%%%%%%%%%%%%%%%%%%%%%%%%%%%%%%%%%%%%%%%%%%%%%%%%%%%%%%%%%%%%%%%%%%%%%%%%%%%%%%%%
\section{General discussion of an Effective Action for the domain wall dynamics}
\label{Sec:general_discussion}
%%%%%%%%%%%%%%%%%%%%%%%%%%%%%%%%%%%%%%%%%%%%%%%%%%%%%%%%%%%%%%%%%%%%%%%%%%%%%%%%%%%%%%

Effective field theory techniques are widely employed in physics to capture the relevant dynamics of 
low-energy degrees of freedom. The core idea is to construct the most general action for these degrees 
of freedom while respecting all the symmetries of the system. However, for this approach to be practical, 
the existence of a small parameter is essential, allowing us to constrain the number of terms in the 
effective action.

This method is valuable because it clarifies which interactions are important and how they scale with the
relevant energy. It typically enables the formulation of an effective theory at a given energy scale, even without knowing the complete underlying theory. Moreover,  when the ``microscopic'' theory is known, it is, in principle, possible to compute the couplings/coefficients that appear in the effective theory. Although this is 
generally a challenging task, we will demonstrate that for the dynamics of domain walls, it is feasible
to compute these couplings explicitly.

In this work, our goal is to derive an effective action governing the dynamics of domain walls. Considering 
the different types of excitations on the walls and their energies, it is natural to first focus on the 
action that captures the dynamics of the zero mode, or in other words, the position of the wall. Specifically, 
we aim to describe the location of the domain wall in spacetime, namely the functions
\beq
x^{\mu}= x^{\mu}(\sigma^a)~,
\eeq
where $\sigma^a$ are the intrinsic coordinates on the worldvolume. From a geometric perspective, this corresponds 
to obtaining the worldvolume of the wall, i.e., the hypersurface that the center of the wall traces through spacetime\footnote{In the $\lambda \phi^4$ model the center of the wall is pretty clearly defined, as $\phi=0$ corresponds to the symmetric point of the kink solution. In other models, there could be some ambiguity about what should be considered the center. We will comment on this when appropriate in the next sections.}.

If we disregard any internal structure of the wall, assuming an infinitely thin domain wall, the simplest 
effective action that comes to mind for its dynamics is the extension of the relativistic particle action, known 
as the Nambu-Goto action \cite{nambu1970lectures,Goto:1971ce}  for extended objects, namely,
\beq
S_{NG} = - \mu \int{d^D\sigma \sqrt{-\gamma}}~;
\eeq
where $\mu$ describes the tension of the wall and we have introduced the induced metric on the worldvolume,
\beqa
\gamma_{ab} = \eta_{\mu \nu} \frac{\partial x^{\mu}}{\partial \sigma^a}\frac{\partial x^{\nu}}{\partial \sigma^b}= \eta_{\mu \nu} e^{\mu}_{a} e^{\nu}_{b}~,\
\eeqa
where we use the notation $e^{\mu}_{a} = \frac{\partial x^{\mu}}{\partial \sigma^a}$ to describe the tangent vectors to the wall. 

To extend the analysis beyond the lowest order and account for the effects of the finite thickness of the 
soliton, additional terms must be included in the effective action. These terms must respect the 
reparametrization invariance of the worldvolume of the wall. Consequently, they should be constructed 
using the worldvolume metric, $\gamma_{ab}$, and the extrinsic curvature,
\beqa
K_{ab} = \nabla_{\mu} n_{\nu} \frac{\partial x^{\mu}}{\partial \sigma^a}\frac{\partial x^{\nu}}{\partial \sigma^b}= \nabla_{\mu} n_{\nu} e^{\mu}_{a} e^{\nu}_{b}~,\
\eeqa
%\bigskip
where we have introduced $n^{\mu}$, a unit 4-vector ($\eta_{\mu \nu} n^{\mu}  n^{\nu} = 1$) normal to the worldvolume, $\eta_{\mu \nu} n^{\mu}  e^{\nu}_{b} = 0$. This procedure indicates that the terms to be considered should be
\beq
S_{HD}= - \mu \int{d^D\sigma \sqrt{-\gamma}\left[1 + c_1 K + c_2 \mathcal{R} + c_3 K^2 + c_4 K_{ab} K^{ab} + ...\right]}~,
\eeq
where $K = \gamma^{ab} K_{ab}$ denotes the trace of the extrinsic curvature and $\mathcal{R}$ is the Ricci
scalar built from the induced metric, $\gamma_{ab}$. 

Naturally, the number of terms in this expression is infinite. However, this effective action can be 
expressed as a series expansion in powers of the small parameter $\frac{l}{L}$, where $l$ represents 
the thickness of the soliton's profile, and $L$ is the characteristic length scale of the extended object\footnote{In a static situation, one can think of this scale as, for example, the radius of a spherical domain wall.}. In order to
do this, we observe that the different geometric terms in the action scale differently with $L$, with the components of the extrinsic curvature scaling as $K_{ab} \sim L^{-1}$ and the Ricci scalar as $\mathcal{R} \sim L^{-2}$. Thus, the action can be systematically rewritten by grouping terms according to their scaling behavior,
\beq
\label{HD-action-group-terms}
S_{HD}= - \mu \int{d^D\sigma \sqrt{-\gamma}\left[1 + c_1 K + \left(c_2 \mathcal{R} + c_3 K^2 + c_4 K_{ab} K^{ab} \right) +
\mathcal{O} (l/L)^3 +...\right]}~,
\eeq
where $c_1\sim \mathcal{O} (l)$, $(c_2,c_3,c_4)\sim \mathcal{O} (l^2)$.
Moreover, in flat space we can use the Gauss-Codazzi equation $\mathcal{R}=K^2 - K_{ab} K^{ab}$, so not all the 
terms at the second order are independent and we can reduce even more the most generic action.

%%%%%%%%%%%%%%%%%%%%%%%%%%%%%%%%%%%%%%%%%%%%%%%%%%%%%%%%%%%%%%%%%%
\subsection{Effective Field Theory for transverse displacements}
%%%%%%%%%%%%%%%%%%%%%%%%%%%%%%%%%%%%%%%%%%%%%%%%%%%%%%%%%%%%%%%%%%

To analyze the conclusions from the effective action terms, we begin by examining the equations of motion
for a domain wall in the simplest possible scenario. Let us consider a nearly flat domain wall with a small, localized perturbation. In this context, it is natural to define the transverse displacement of the wall 
from its fully straight configuration as the single dynamical degree of freedom. This approach suggests
aligning the intrinsic worldsheet coordinates with the Minkowski spacetime coordinates, except for a single 
coordinate representing the wall's displacement. Thus, in this gauge, the worldsheet can be parametrized 
as follows:
\beq
x^{\mu} = \left\{\sigma^1, \ldots, \sigma^D, \psi(\sigma^a)\right\}\,.
\eeq

Using this parametrization and plugging it in Eq. (\ref{HD-action-group-terms}) we will obtain an action of 
the form
\beq
S_{\psi} = - \mu \int{d^D\sigma \sqrt{1+ \eta^{ab} \partial_a \psi \partial_b \psi}\left[1+ \mathcal{O}(\partial^2 \psi) +...\right]}\,.
\eeq

We can now think about this action as a field theory model in $D$ spacetime dimensions for a 
scalar field $\psi$. At the lowest order (just considering the NG term), the action describes a non-linear
theory with all orders of first derivatives\footnote{In \cite{Blanco-Pillado:2022rad} it was shown, using the original scalar field theory model, that indeed the collision of waves in the $\psi$ field interact (at the lowest order) following this non-linear Born-Infeld action for $\psi$.}. Furthermore, our analysis indicates that incorporating higher curvature terms in the original action for the wall results in a higher-derivative action for the scalar field $\psi$ describing the wall’s displacement. Here we have denoted all these terms collectively by $\mathcal{O}(\partial^2 \psi)$. However, such higher-derivative actions are known to introduce problematic equations of motion, often accompanied by ghost states and nonphysical solutions. Given that our effective action originates from a well-behaved field theory,  it is hard to see how one can admit such pathological solutions. Therefore, it would be desirable to find a framework that allows us to construct actions that avoid all these issues.

This type of higher-derivative action has been extensively discussed in the literature, and strategies exist for identifying effective theories with well-defined evolution. Specifically, \cite{Nicolis:2008in} demonstrated that the most general theory with stable dynamics can be constructed by limiting the action to a specific subset of terms. These are known as Galileon theories, in which the scalar field satisfies a specific symmetry.

Furthermore, \cite{deRham:2010eu} established a very important connection between Galileon theories and the degrees of freedom of branes in a higher-dimensional background. Here, the special symmetry of Galileon theories is understood as the non-linear realization of the original higher-dimensional theory’s Poincaré symmetry. This work also demonstrated that for the effective theory of brane displacements to be a valid Galileon theory, certain terms in the brane’s effective action must be restricted.

Applying this approach to our case, we conclude that further restrictions are necessary for the general action given in Eq. (\ref{HD-action-group-terms}). Specifically, we determine that the most general action free from higher-derivative equations of motion can be expressed as follows:
\beq
S= - \mu \int{d^D\sigma \sqrt{-\gamma}\left[1 + C_1 K + C_2 \mathcal{R} + C_4 \mathcal{K}_{GB}\right]}~,
\eeq
where 
\beq
\mathcal{K}_{GB} = - \frac{2}{3} K_{a b}^3 + K K_{a b}^2 - \frac{1}{3}K^3 - 2\left(\mathcal{R}_{ab} - \frac{1}{2}
\mathcal{R} \gamma_{ab} \right) K^{ab}
\label{GaussBonnetterm}
\eeq
describes a brane term inherited from a Gauss-Bonnet term (see \cite{deRham:2010eu}).

This approach is interesting because it significantly simplifies the effective action, both in terms of 
the limited possible lowest-order terms and also by clarifying that the action is truncated at a specific 
order. As a result, the analysis requires calculating only a limited set of coefficients derived from 
the higher-dimensional field theory. We turn to this computation next.

%%%%%%%%%%%%%%%%%%%%%%%%%%%%%%%%%%%%%%%%%%%%%%%%%%%%%%%%%%%%%%%%%%%%%%%%%%%%%%%%%%%%%%
\section{Deriving the Effective Action from an underlying Scalar Field Theory}
\label{Sec:deriving_action}
%%%%%%%%%%%%%%%%%%%%%%%%%%%%%%%%%%%%%%%%%%%%%%%%%%%%%%%%%%%%%%%%%%%%%%%%%%%%%%%%%%%%%%

In order to proceed we need to be able to compute the coefficients of the effective action
we presented earlier directly from the field theory model described by Eq. (\ref{FT-action}). For this, we will
adopt the methodology outlined in \cite{LinLowe83,Zia:1985gt}, where a comparable calculation 
was performed for the effective theory of Euclidean interfaces. We change the computation to adjust
the calculation to our current analysis in a Lorentzian framework whenever necessary.

We start by considering a domain wall of arbitrary shape determined by the surface $z=\psi(y^a)$ in flat space coordinates given by $x^\mu =(y^a,z)$\footnote{With a small abuse of notation, we will denote both a Lorentzian vector and its components with the same symbol, e.g., $x^\mu$}. On the other hand, let us now define a new curvilinear coordinate system for flat spacetime given by $\zeta^\mu=(v^a,u)$. The change of coordinates between these two systems
can be expressed by
\begin{equation}
   x^\mu\equiv (y^a,z)=\xi^\mu(v)+u~n^\mu(v)~,
\end{equation}
where $\xi^a=v^a$, $\xi^{d+1}=\psi(v)$, and $n^\mu$ is the $D$-dimensional unit vector normal to the wall at each point in $v$:
\begin{equation}
    n^\mu=\frac{(-\nabla_a\psi(v),1)}{\sqrt{- \gamma(v)}}~,\quad~~~~~~ \gamma=-(1+ \eta^{ab}\nabla_a\psi\nabla_b \psi)\,,
    \label{metric_worldsheet}
\end{equation}
where we have denoted $\nabla_ a=\frac{\partial}{\partial v^a}$. We will denote partial derivatives with respect to the flat coordinate $y^a$ by $\partial_a$. 

\begin{figure}
    \centering
    \includegraphics[width=0.8\linewidth]{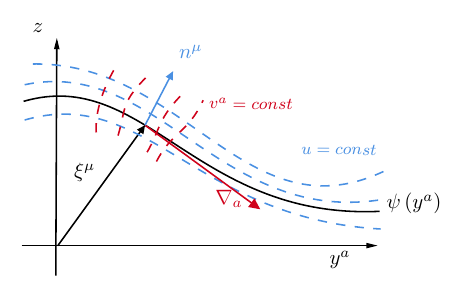}
    \caption{Adapted normal coordinates to the wall. Note that one of the $y^a$ is timelike, so the figure represents a spacetime diagram.}
    \label{fig:coord_sets}
\end{figure}

Note that, in the new coordinate system, the surface with $u=0$ corresponds to the
position of the wall in spacetime which is parametrized by the $v^a$ coordinates. From the field theory point of view, we will define the center of the wall as the locus of zeros of the scalar field $\phi$ in the $\lambda \phi^4$ model although this definition is somewhat arbitrary for non-symmetric domain walls\footnote{Indeed, the level sets of $\phi$ give a foliation of the spacetime manifold into a family of $D$-dimensional submanifolds, each of which can be defined to be the worldsheet of the wall. Here we define the center to be the set with $\phi =0$ in the $\lambda \phi^4$ model but one could instead define the center  as the level set corresponding to the field values in which the static energy density is maximum. This may be more physically justified in asymmetric kink models.}. Going into the
perpendicular direction along the vector $n^{\mu}$ we can cover the vicinity of the domain wall in 
spacetime at least for a while\footnote{We will comment on the limitation of this coordinate system
later on in the text.} (see \cref{fig:coord_sets} for more clarification of these two coordinate systems).

The Jacobian of the transformation is $|\det e^\alpha_\beta|$, which we state here for future reference:
\begin{equation}
e_\alpha^\mu = \frac{\partial x^\mu}{\partial \zeta^\alpha} = 
\begin{cases} 
\nabla_a \xi^\mu + u \nabla_a n^\mu, & \alpha = a~, \\ 
n^\mu, & \alpha = D ~.
\end{cases}
\end{equation}

We can now express the form of the field theory action given by Eq. (\ref{FT-action}) in the coordinate 
system adapted to the domain wall. A direct computation shows that the result is given by
\beq
S=-\frac 12 \int d^{D}v\int du |\det e^\alpha_\beta|~\left[e^\mu_\alpha\partial_\mu\phi\partial^\lambda\phi e_\lambda^\alpha+V(\phi)\right].
\eeq

Moreover, since the coordinate system we are using is defined such that the center of the domain wall profile is positioned at $u=0$, we will assume that surfaces of constant $u$ near that value also represent constant field surfaces. Hence we have
\begin{equation} e^\mu_\alpha\partial_\mu\phi\partial^\lambda\phi e_\lambda^\alpha=(\partial_u\phi)^2 n_\mu n^\mu=(\phi'_K)^2.
\end{equation}

On the other hand, 
\begin{equation}
|\det e^\alpha_\beta|=\sqrt{|\det G_{\mu\nu}|}~,
\label{eqT1}
\end{equation}
where $G_{\mu\nu}$ is just the flat metric written in the new coordinates, which takes the form
\begin{equation}
G_{\mu\nu}=\left(
\begin{matrix}
G_{a b} & 0\\
0 & 1
\end{matrix}\right)~~~~~~;~~~~~~G_{ab}=(\delta_a^c-uK^c_a)\gamma_{cd}(\delta_b^d-uK^d_b)~,
\label{eqT2}
\end{equation}
where $\gamma_{ab}$ and  $K_a^b$ denote the induced metric on the wall and the extrinsic curvature respectively. From (\ref{eqT1}) and (\ref{eqT2}), and the fact that the static equations of motion for the soliton satisfy $\phi_K'=\sqrt{V(\phi_K)}$, we have
\begin{equation}
    S=-\int d^Dv\int du\sqrt{-\gamma(v)}~|\det[I-u{\bf K}(v)]|~(\phi_K'(u))^2,
    \label{General_action}
\end{equation}
where we denote by $\bf K$ the $D\times D$ matrix of extrinsic curvature, whose matrix elements are given by $K^a_b$.
The above expression is interesting because it implies that there will only be a finite number of curvature correction terms in the effective action, and furthermore it also predicts both the exact terms and their associated coefficients. Indeed, the determinant can be expanded as \cite{Curtright:2020cta}
\begin{equation}
    \det[I-u {\bf K}]=\sum\limits_{n=0}^D\frac{(-u)^n}{n!}\mathcal{E}_n({\bf K})~,
\end{equation}
where 
\begin{equation}
    \mathcal{E}_n({\bf K})=\delta^{a_1\cdots a_n}_{b_1\cdots b_n}K^{b_1}_{a_1}\cdots K^{b_n}_{a_n}~,
\end{equation}
so that the effective action, after integrating over the transversal coordinate $u$, becomes
\begin{equation}
    S_{\rm eff}=-\int d^Dv\sqrt{-\gamma(v)}~ \sum\limits_{n=0}^D \tilde c_n\mathcal{E}_n({\bf K})~,
\label{effective_Goldstone}
\end{equation}
where the coupling constants $\tilde c_n$ are defined as
\begin{equation}
    \tilde c_n= \int \frac{(-u)^n}{n!}(\phi'(u))^2 du~.
    \label{couplings}
\end{equation}
We see that the number of terms appearing in the effective action depends on the number of dimensions of the ambient space $D$. In particular, the first terms are
\beqa
    \mathcal{E}_0({\bf K})&=&1~,\\
    \mathcal{E}_1({\bf K})&=&K_a^a\equiv K~,\\
    \mathcal{E}_2({\bf K})&=&\frac{1}{2}\left(K^2-K_{a}^b K_{b}^a\right)\equiv \frac 12 \mathcal{R}~,\\
    \mathcal{E}_3({\bf K})&=&\frac{1}{3!}\left(K^3-3KK_{a}^bK^{a}_b+2K_{a}^bK^{c}_bK^a_c\right)~,\\
    \vdots \notag\\[2mm]
\mathcal{E}_D({\bf K})&=&\det(K^a_{b})~.
\eeqa

This derivation demonstrates that the effective action contains only a finite number of terms for a given worldsheet dimension $D$. Moreover, these terms match those introduced in the previous section and correspond to the \emph{DBI-Galileon} terms \cite{deRham:2010eu}. This statement is trivial for the first two terms, but for the Gauss-Bonnet term one must take into account that, for a metric of the form \eqref{metric_worldsheet}, the Riemann curvature tensor can be written in terms of the extrinsic curvature \cite{Zia:1985gt}:
\begin{equation}
    R_{\mu\nu\alpha\beta}=K_{\mu\alpha}K_{\nu\beta}-K_{\mu\beta}K_{\nu\alpha}.
\end{equation}
Hence, by rewriting the Einstein tensor appearing in \cref{GaussBonnetterm} in terms of the extrinsic curvature, we find
\begin{equation}
    \mathcal{E}_3(\mathbf{K})=\frac{1}{4}\mathcal{K}_{\rm GB}
\end{equation}
as expected.

Being constructed from Galileon terms, the equations of motion derived from action \eqref{effective_Goldstone} are second-order and free from pathologies. More importantly, the procedure detailed above provides a direct route to calculate the numerical coefficients of the action from the underlying field theory. With these coefficients in hand, the effective theory is fully equipped to make concrete predictions about the dynamics of the domain wall solitons present in the theory.

%%%%%%%%%%%%%%%%%%%%%%%%%%%%%%%%%%%%%%%%%%%%%%%%%%%%%%%%%%%%%%%%%%%%%%%%%%%%%%%%%%%%%%
\section{Comparing with Field Theory: Collapsing Closed Walls}
\label{sec:numerics_effective}
%%%%%%%%%%%%%%%%%%%%%%%%%%%%%%%%%%%%%%%%%%%%%%%%%%%%%%%%%%%%%%%%%%%%%%%%%%%%%%%%%%%%%%

In the previous sections, we derived the effective action governing the dynamics of
the position of the domain walls in field theory including higher curvature corrections. This action 
reveals a few intriguing facts. First, the coefficients
in front of each of the terms depend on the particular field theory model at hand. Second, the 
dimensionality of the system dictates the number of terms that appear. And lastly, some terms may 
be topological, therefore not contributing to the equations of motion we derive from these effective 
actions. 

In the following sections, we will explore various soliton solutions in different field theory models 
across multiple dimensions, examining their dynamics to assess the precision and relevance of the 
effective actions derived earlier. For comparison, we will also simulate the domain wall evolution 
using the full field theory dynamics within a lattice field theory framework (details of this numerical 
lattice field theory implementation can be found in appendix \ref{Numerical-FT}).

%%%%%%%%%%%%%%%%%%%%%%%%%%%%%%%%%%%%%%%%%%%
\subsection{2+1 Dimensions}
%%%%%%%%%%%%%%%%%%%%%%%%%%%%%%%%%%%%%%%%%%%

Field theory domain walls in $2+1$ dimensions are better described as domain wall-strings,
since they correspond to line-like defects in $2$ spatial dimensions evolving in time.
According to the description in the previous sections, their effective action
should be well approximated by

\beq
\label{2+1-Eff}
S= - \mu \int{d^2\sigma \sqrt{-\gamma}\left[1 + C_1 K \right]}~.
\eeq

In principle, one might consider adding a term proportional to $\mathcal{R}$; however, in a 1+1-dimensional 
worldsheet, this term is purely topological and thus it would not affect the equations of motion. Furthermore, additional higher order terms in \eqref{effective_Goldstone} vanish identically in two dimensions, hence will not contribute to the dynamics.

To validate this action, we will analyze a simple setup: a cylindrically symmetric domain wall-string 
initially at rest. It is clear that given this initial condition the tension of the defect will cause the domain wall ring to shrink and finally collapse. Here, our focus is on comparing the detailed trajectory of the soliton's center observed in a full field theory simulation with the trajectory predicted by this effective action.

In order to compute the evolution of the ring we parametrize the worldsheet in cylindrical coordinates as
\beq
x^{\mu} = (t, \rho(t), \theta)~,
\label{coords2+1}
\eeq
so the equation of motion that controls the radius of the ring is\footnote{See the derivation in the appendix \ref{app:details}.}
\beq
\label{rho-equation}
\ddot \rho = -\frac{\left(1-\dot \rho^2\right)^{3/2}}{2 C_1+\rho \sqrt{1-\dot\rho^2}}~.
\eeq

Note that as we described in previous sections the coefficient of the trace of the
extrinsic curvature can be computed using the expression
\beq
C_1 = \frac{\tilde c_1}{\tilde c_0} = - \frac{1}{\mu} \int{u ~(\phi'(u))^2 du}~,
\label{eq:C1}
\eeq
where $\phi(u)$ denotes the static solution for our domain wall profile.
Therefore, it is clear that the sign of this coefficient depends on the degree of
asymmetry of the soliton solution with respect to the center of the object, the point
where $u=0$.

In the following subsections we will investigate the effect of this asymmetry on
the domain wall dynamics and how well it can be described by the effective 
action of the form given by Eq. (\ref{2+1-Eff}).

%%%%%%%%%%%%%%%%%%%%%%%%%%%%%%%%%%%%%
\subsubsection{\texorpdfstring{$\lambda \phi^4$}{lphi4}}
%%%%%%%%%%%%%%%%%%%%%%%%%%%%%%%%%%%%%

The soliton solution for the $\lambda \phi^4$ model is provided explicitly in Eq. (\ref{kink-f4}), 
and it clearly satisfies the symmetry condition, $\phi(z) = - \phi(-z)$, which is inherited from 
the symmetry of the Lagrangian. Consequently, examining the form of the coefficient $C_1$ 
immediately shows that it must be zero. The resulting equation of motion in Eq. (\ref{rho-equation}) with 
$C_1=0$ is sufficiently simple to allow integration, yielding a solution of the form
\beq
\label{rho-equation-2}
\rho_{NG}(t) = \rho_0 \cos\left(\frac{t}{\rho_0}\right)~,
\eeq
where $\rho_0$ denotes the initial radius of the domain wall string at rest at the initial time,
$t=0$.

\begin{figure}
    \centering
    \includegraphics[width=0.8\linewidth]{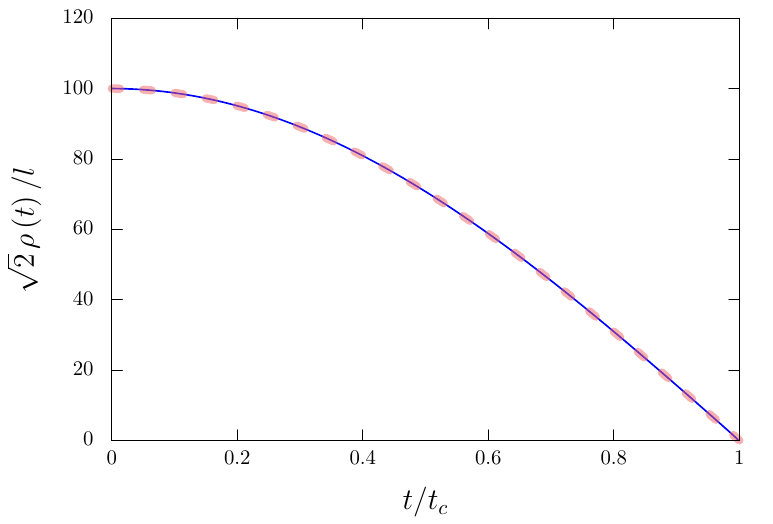}
    \caption{Radius of the circular $\lambda\phi^{4}$ wall as a function of time. The solid blue curve corresponds to the field theory simulation, and the thick pink dashes represent the effective solution (\ref{rho-equation-2}). Time is displayed in units of the collapse time $t_{c}=\pi\rho_{0}/2$, with $\rho_{0}=\left(100/\sqrt{2}\right)l$, and the radius is given in units of the length scale $l=\sqrt{2/\lambda}\,\eta^{-1}$.}
    \label{fig:circular-wall}
\end{figure}

In \cref{fig:circular-wall}, we compare the evolution of the soliton center obtained from the full field theory 
simulation to the analytic solution derived above. Notably, in this case, there are no corrections 
to the lowest order, the Nambu-Goto action, as all higher-order terms vanish. Nonetheless, we include 
the comparison of the different numerical schemes as a benchmark to illustrate the level of precision 
achievable in our analysis. For most of the time, except for the final stage of the collapse, the deviations 
with respect to the Nambu-Goto solution are less than $0.03\%$.

Here we would like give a short explanation of the rationale behind our choice of initial conditions. For a meaningful comparison between the field theory solutions obtained via lattice-based numerical evolution and the predictions of the worldsheet effective action, the initial field configuration must closely approximate an exact solution. Otherwise, discrepancies in evolution could stem from errors in the initial setup rather than from misidentified terms in the effective action. To minimize such uncertainties, we use a cylindrical configuration starting from rest with a large initial radius compared to the thickness of the soliton profile. This approach reduces errors in the field theory initial condition, as it relies solely on the uncorrected static soliton profile strictly only valid for a flat domain wall. A side effect of this procedure is that the corrections we would like to identify in the dynamics of the wall are very small, so we need to introduce enough points in our simulations to achieve the accuracy necessary to correctly capture small deviations due to higher curvature corrections. A detailed  explanation of the initial conditions, along with the methodology for extracting the relevant data from field theory simulations, is provided in the appendix \ref{Numerical-FT}.

One can think of several different numerical experiments to probe the validity of our effective actions apart from the one presented here. A compelling example could involve investigating the collision of domain wall wiggles. Such explorations were undertaken in \cite{Blanco-Pillado:2022rad} primarily to validate the accuracy of Nambu-Goto dynamics and quantify radiation produced during the scattering process. In principle, these simulations might reveal the influence of higher-derivative corrections. However, as demonstrated in this work, no such corrections are expected in the $\lambda \phi^4$ model. This insight likely explains the observations in \cite{Blanco-Pillado:2022rad}: the simulations transitioned seamlessly from NG dynamics to non-perturbative radiation emission, bypassing any intermediate regime where additional terms in the effective action might play a role.

%%%%%%%%%%%%%%%%%%%%%%%%%%%%%%%%%%%%%
\subsubsection{\texorpdfstring{$\lambda \phi^6$}{lphi6}}
%%%%%%%%%%%%%%%%%%%%%%%%%%%%%%%%%%%%%

We can easily generate a non-vanishing coefficient for the trace of the extrinsic
curvature term by changing our field theory slightly. One of the simplest
models that generates this asymmetry but is still analytically tractable is the
$\phi^6$ model \cite{Lohe:1979mh}, namely,
\beq
S = \int{ d^{D+1}x \left[-\frac{1}{2} \partial_{\mu} \phi \partial^{\mu} \phi - \frac{\lambda}{4} \left(\phi^2 - \eta^2\right)^2 \phi^2\right]}~,
\eeq
with static solutions of the form
\begin{equation}
    \Phi_K^{\pm}(x,t)=\pm\eta\sqrt{\frac{\tanh{(\ell^{-1}x)}+1}{2}}\,,
\end{equation}
where the characteristic width is now defined as $\ell =\sqrt{\frac{2}{\lambda}}\eta^{-2}$.
\begin{figure}
    \centering
    \includegraphics[width=0.7\linewidth]{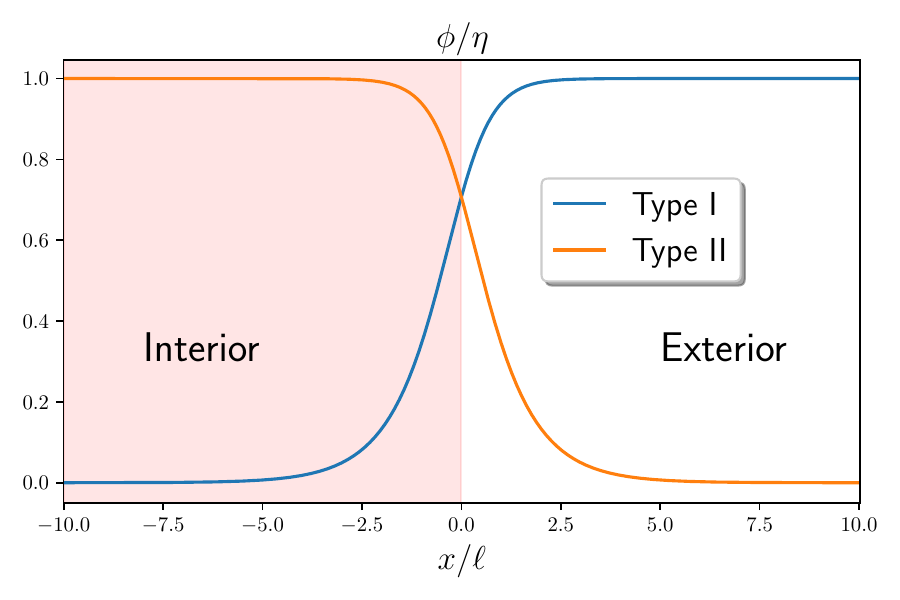}
    \caption{The two inequivalent types of closed domain wall in the $\phi^6$ model. Note they are not symmetric under inversion about their center.}
    \label{fig:dwtypes}
\end{figure}

In this modified Lagrangian, the soliton solution exhibits an asymmetry under central inversion\footnote{Technically, as opposed to the case of center-symmetric kinks,  there is not a unique notion of 'center' for asymmetric kinks. Formally, in the procedure we have developed, the resulting effective action does depend on the (arbitrary) choice of a center. However, this choice is equivalent to choosing a (hyper)surface of constant field values at a given initial instant. Then, the effective action will reproduce the dynamics of the worldsheet associated to the evolution of such hypersurface.}, resulting in a non-zero 
$C_1$  coefficient. Indeed, the value of this coefficient in the $\phi^6$ model is $C_1=\pm(\sqrt{2\lambda}\eta^2)^{-1}\equiv \pm2\ell$, depending on the orientation of the wall. This allows for two distinct types of cylindrical walls, distinguished by which 
vacuum occupies the interior and which occupies the exterior. We refer to these configurations as 
Type-I and Type-II, as shown in \cref{fig:dwtypes}.

\begin{figure}[hbt!]
     \centering
     \begin{subfigure}[b]{0.45\textwidth}
         \centering
         \hspace*{-1.2cm}\includegraphics[width=1.2\textwidth]{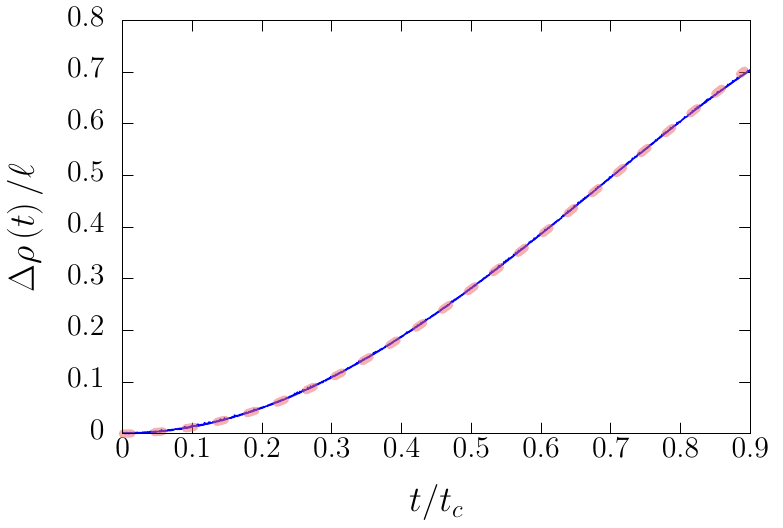}
         \caption{Type I domain wall}
         \label{fig:TypeI}
     \end{subfigure}
     \hfill
     \begin{subfigure}[b]{0.45\textwidth}
         \centering
         \hspace*{-1.0cm}\includegraphics[width=1.2\textwidth]{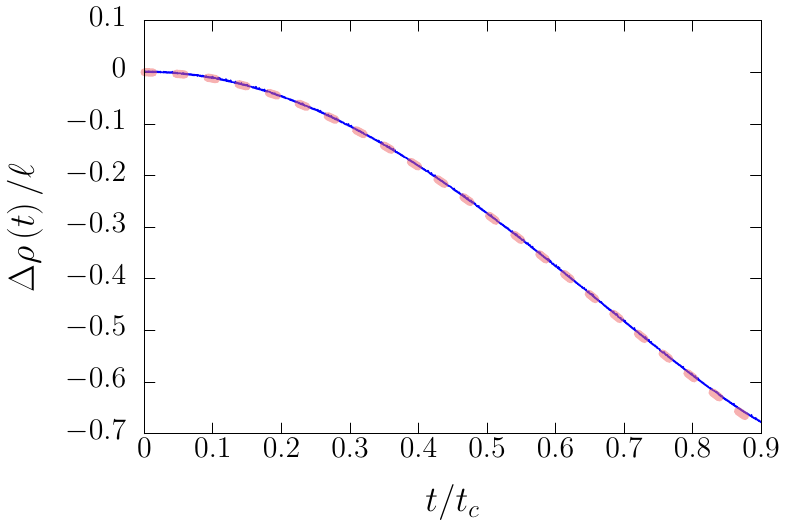}
         \caption{Type II domain wall}
         \label{fig:TypeII}
     \end{subfigure}
        \caption{$\rho_{\text{NG}}(t)-\rho(t)$ curves for the two types of domain walls in the $\phi^6$ model. The solid blue curves represent the result in field theory, and the dashed pink ones represent the prediction given by the effective model \eqref{2+1-Eff}. As in \cref{fig:circular-wall}, $t_{c}=\pi\rho_{0}/2$, with $\rho_{0}=100\ell$. As mentioned in the main text, $\ell=\sqrt{2/\lambda}\,\eta^{-2}$.}
        \label{fig:phi6_comp}
\end{figure}

We show in \cref{fig:phi6_comp} the comparison between the numerical evolution of these two 
configurations and the predictions from the effective action obtained by a numerical integration
of Eq. (\ref{rho-equation}). In order to make a meaningful comparison, we show the deviation of the
position of the wall with respect to the case without higher curvature correction, the pure Nambu-Goto
dynamics. As expected, these deviations for the two types of configurations differ 
due to the vacuum asymmetry, which is captured in the effective action by the sign change of the 
$C_1$ coefficient.

The results in this section confirm that the curvature corrections employed to compute the wall's dynamics 
are indeed correct. Moreover, the method used to derive their coefficients from field theory accurately 
reproduces the observed evolution of the walls.

%%%%%%%%%%%%%%%%%%%%%%%%%%%%%%%%%%%%%%%%%%%
\subsection{3+1 Dimensions}
%%%%%%%%%%%%%%%%%%%%%%%%%%%%%%%%%%%%%%%%%%%

We now extend our analysis to $3+1$ dimensions. An interesting aspect of this transition is 
that the additional dimension allows a new term in the effective action to become dynamical 
and play a significant role in the equations of motion. Specifically, in this case, the most 
general form of the effective action is given by
\beq
\label{3+1-Eff}
S= - \mu \int{d^D\sigma \sqrt{-\gamma}\left[1 + C_1 K + C_2 \mathcal{R}\right]}~,
\eeq
where $C_1$ is given in \cref{eq:C1} and 
\begin{equation}
    C_2 = \frac{\tilde c_2}{\tilde c_0} =  \frac{1}{2\mu} \int{u^2 ~(\phi'(u))^2 du}~.
\end{equation}

Following the approach in the previous section, we now analyze the behavior of spherical domain 
wall bubbles. These spherical domain walls also undergo collapse within a finite time, but the 
exact trajectories will vary depending on the specific field theory model under consideration. Such 
differences directly influence the effective action, particularly through the $C_i$ coefficients. To 
test this concept, we evolve these configurations by solving the field theory equations of motion 
and then compare the results with the evolution equation for the bubble radius, which in this case 
takes the form (see \cref{app:details} for details of its derivation)
\begin{equation}
    \ddot R=\frac{2 \left(\dot R^2-1\right)^2 \left(C_1+R \sqrt{1-\dot R^2}\right)}{R \left(\dot R^2-1\right) \left(4 C_1+R \sqrt{1-\dot R^2}\right)-6 C_2 \sqrt{1-\dot R^2}}~~.
    \label{EOM}
\end{equation}

In the previous section, we validated the role of the $K$ term in the effective action for the
$2+1$ dimensional case. Here, we aim to assess the necessity of the $\mathcal{R}$ term in $3+1$ dimensions. 
Including both terms, however, complicates numerical comparisons. Fortunately, insights into how 
coefficients depend on the model offer a practical solution. As discussed, in a $\lambda \phi^4$ model, 
the $C_1$ coefficient vanishes, making this an ideal case to isolate and examine the presence of a 
higher-order term—in this case, the $\mathcal{R}$ term. In the $\lambda\phi^4$ model, we have 
\beq
C_2=\frac{(\pi^2-6)}{12 \lambda\eta^2}\equiv \frac{(\pi^2-6)}{24}l^2~.
\eeq

\begin{figure}
    \centering
    \includegraphics[width=0.8\linewidth]{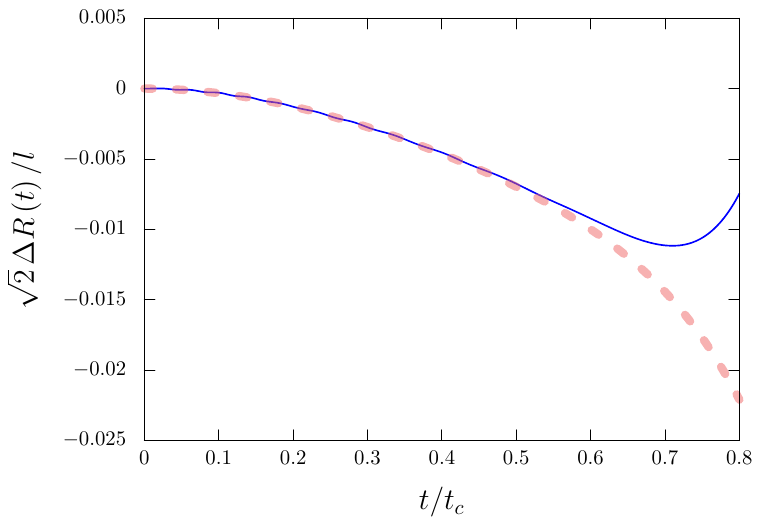}
    \caption{$R_{\textnormal{NG}}(t)-R(t)$ for the $\lambda\phi^{4}$ wall in $3+1$ dimensions. The solid blue curve shows the result obtained in a field theory simulation, while the dashed pink one corresponds to the solution to the effective equation (\ref{EOM}). Time is displayed in units of the collapse time $t_{c}\approx 92.7\,l$, which can be obtained from (\ref{NG 3plus1}) for an initial radius $R_{0}=\left(100/\sqrt{2}\right)l$.}
    \label{fig:spherical-wall-1}
\end{figure}

This approach provides a clear pathway to test for the additional term’s relevance. This is 
exactly what we do in this section. We show in \cref{fig:spherical-wall-1} a comparison of 
the evolution of a spherical domain wall, initially at rest, as predicted by the $\lambda \phi^4$ scalar 
field theory and our effective action given by Eq. (\ref{3+1-Eff}). To facilitate a direct visual comparison, we plot $\Delta R(t)=R_{\textnormal{NG}}(t)-R(t)$
that captures the deviation of the wall's radius evolution from the pure Nambu-Goto prediction\footnote{We give in the \cref{app:details} the analytic solution of the spherical wall collapse written in terms of Jacobi elliptic functions.} for both methods.

The numerical solutions for both the field theory and the bubble radius from our effective theory show remarkable
agreement in \cref{fig:spherical-wall-1} over a substantial period relative to the total collapse time. However,
eventually, the prediction obtained from the effective action stops being a good approximation. 
This discrepancy highlights an important factor: the field theory model contains many more modes
beyond the single mode tracking the soliton's position. As discussed earlier, additional soliton
excitations can be found by studying small perturbations of the soliton, and so one can imagine
that some of these modes could  become significant as collapse approaches. The numerical field theory
solution suggests that this is precisely what happens in our simulations as we get
closer to the collapse. It would therefore be desirable to find an effective action that
incorporates the effects of these other modes and their interaction with the
position of the domain wall. We will address this issue in section \ref{sec:boundstate_eff}.

Before that, let us first examine another aspect of the solution that remains accessible using the current formulation of the effective action for the soliton's position, namely the corrections to the
soliton profile due to curvature effects. This is what we turn to in the next section.

%%%%%%%%%%%%%%%%%%%%%%%%%%%%%%%%%%%%%%%%%%%%%%%%%%%%%%%%%
\section{Curvature Corrections to the Soliton's Profile}
\label{Sec:profile_corrections}
%%%%%%%%%%%%%%%%%%%%%%%%%%%%%%%%%%%%%%%%%%%%%%%%%%%%%%%%%

The profile of the field along the co-dimension of the wall gets affected in general by its velocity, acceleration and spatial curvature. While one can get rid of velocity corrections by performing a Lorentz boost to the instantaneous rest frame at each point (i.e. by choosing the appropriate adapted coordinate system), other corrections can be important in 
particular situations.

\begin{figure}[hbt!]
    \centering
    \includegraphics[width=0.495\linewidth]{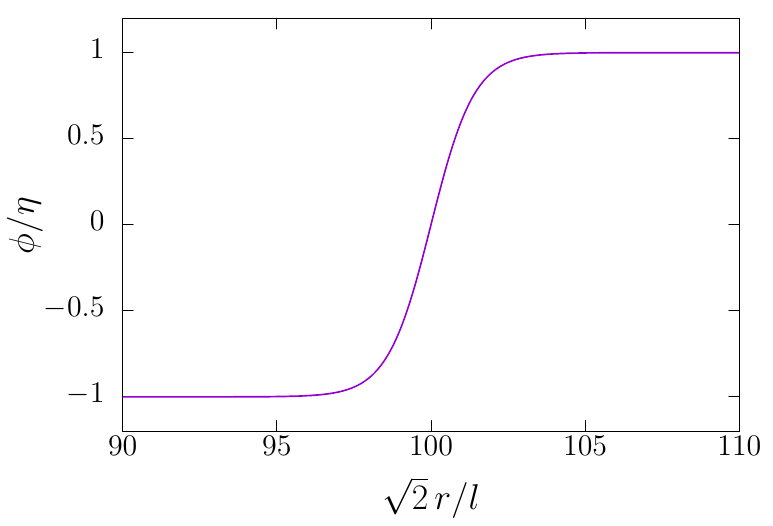}
    \includegraphics[width=0.495\linewidth]{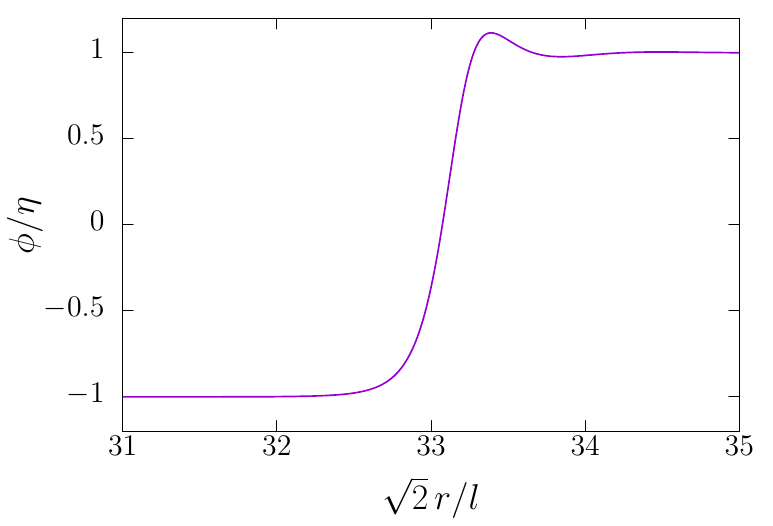}
    \caption{Field profile $\phi\left(r,t\right)$ at $t=0$ (left) and $t\approx 69.3\,l$ (right) for the $\lambda\phi^{4}$ wall in $3+1$ dimensions as a function of the radial coordinate $r$. Apart from Lorentz contraction (note the difference in the scale of the horizontal axis), the profile gets curvature corrections showing up as little bumps around the core of the wall.}
    \label{fig:field-profile}
\end{figure}

In fact, some of these profile distortions have been previously noticed in numerical field theory simulations for a domain wall collapse \cite{Widrow:1989vj}. Here we reproduce in \cref{fig:field-profile} this field profile in one of our simulations for a spherical wall collapsing in $3+1$ dimensions where one can easily
identify the presence of these corrections in the form of a bump in the profile.

The purpose of this section is to show, using the simple assumptions regarding the form of the soliton profile in the
adapted coordinates, how one can obtain the exact shape of these deformations. Therefore, these
profile distortions have nothing to do with radiation but are only the result of a transformation
between the two coordinate systems used to describe the soliton evolution in spacetime.

Our starting point to show this is the expression found in \cref{u-expansion} relating the normal 
parameter perpendicular to the worldsheet (the $u$ parameter) and the original flat space coordinates, namely,
\begin{equation}
\label{expansion-of-u}
    u=\frac{z-\psi}{\sqrt{1+ \eta^{ab} \partial_a\psi\partial_b\psi}}-\frac{1}{2}\frac{(z-\psi)^2}{(1+\eta^{ab} \partial_a\psi\partial_b\psi)^{5/2}} \eta^{ab} \eta^{cd}\partial_a\psi\partial_c\psi\partial_{bd}\psi+\order{(z-\psi)^3}\,.
\end{equation}

Our main assumption here is that at the lowest order of perturbation theory the solution of 
the profile for the field is purely given in terms of $u$. This means that a modification of 
the position of the soliton with respect to the $z$ coordinate and parametrized by $\psi$ would 
be encoded in the field by obtaining the field profile of the form $\phi_K[u(\psi)]$.

One can try to understand the relevance of these two terms in Eq. (\ref{expansion-of-u}) by 
thinking of a purely spatial deformation of the wall with respect to the cartesian coordinates 
in the original flat spacetime. It is clear that a rigid rotation of the soliton solution is 
still a solution of the full non-linear equations of motion.
Furthermore, this pure rotation would not introduce second derivatives in the field $\psi$ with
respect to the internal coordinates, so the second order correction should be zero in this case.
Looking at the first term one can see that indeed this term is sufficient to describe the exact solution 
of a pure soliton rotation\footnote{This was already pointed out in \cite{Blanco-Pillado:2022rad}, where it was shown that an ansatz for the scalar field profile that considers this term would be able to accommodate several
known solutions, like a pure boost or a rotation.}.

Let us now think of a soliton with some curvature in the spatial dimensions. It is clear that
any effect on the soliton's profile due to this can not be accounted for with the first term since this
curvature would depend on second derivatives of $\psi$. It is therefore clear that this second term is 
the one introducing a correction to the surfaces of constant field due to the curvature of the soliton.

In a general situation we can use the expression above to write down the field configuration in the
original flat spacetime frame by just re-writing $\phi_K[u]$ in terms of the original coordinates, namely\footnote{Here we simplify the notation and omit the flat metric on the brane but we note that all the internal indexes contracted must be thought to be contracted with the flat metric $\eta^{ab}$ (namely, when we write here $\partial^a\psi\partial_a\psi$, for example, we really mean $\eta^{ab} \partial_a\psi\partial_b\psi$ and so on).}

\begin{equation}
    \phi(z,y^a)=\phi_K[u]\approx \phi_K\qty(\frac{z-\psi}{\sqrt{1+\partial_a\psi\partial^a\psi}})-\frac{(z-\psi)^2}{2}\frac{\partial^a\psi\partial^b\psi\partial_{ab}\psi}{(1+\partial^a\psi\partial_a\psi)^{5/2}}\phi_K'\qty(\frac{z-\psi}{\sqrt{1+\partial_a\psi\partial^a\psi}})+\cdots
\label{phi-correction}
\end{equation}

One interesting point to notice is that the deformations presented here greatly simplify in the case where $\partial^a\psi\partial_a\psi = 0$ where the field simply becomes
\begin{equation}
    \phi(z,y^a) \approx \phi_K\qty(z-\psi)~.
\end{equation}
 
These configurations are well known and in fact are solutions of the full non-linear
equations of motion and represent traveling waves moving at the speed of light along the
longitudinal directions of the wall \cite{Vachaspati:1990sk}. It is nice to see that our
approximate solutions reduce to these known exact solutions in the correct limit.

Note that, in our approach, the correction to the kink profile comes directly from the inverse coordinate transformation \eqref{expansion-of-u}, and it does not have to satisfy a particular equation, as it happens 
for example in other approaches \cite{Gregory:1989gg}. Further corrections can also be computed systematically, by grinding through the process of calculating the next order terms in 
Eq. (\ref{expansion-of-u}) (for more details see the computation in \cref{u-expansion}).

Finally, we would like to use our numerical simulations in field theory to investigate whether 
the distortions observed in the soliton profile during collapse are primarily artifacts of our coordinate frame. By employing the coordinate transformation discussed earlier and assuming that the soliton's 
profile remains fixed in its rest frame, we can compute the profile as it would appear from the 
laboratory frame, which is the frame of our simulation.

In the case of a collapsing spherical wall, we can choose the radial direction as the transverse co-dimension in the lab frame. Due to spherical symmetry, the position of the wall does not depend on additional spatial coordinates, and it will just be given as a function of time, $\psi(y^a)\equiv R(t)$. Then, we just need to rewrite the expression for the field presented in Eq. (\ref{phi-correction}) in terms of spherical coordinates,

\begin{equation}
    \phi(r,t)\approx\phi_K\qty(\frac{r-R}{\sqrt{1-\dot{R}^2}})-A(t)\qty(\frac{r-R}{\sqrt{1-\dot{R}^2}})^2\phi_K'\qty(\frac{r-R}{\sqrt{1-\dot{R}^2}})+\cdots,
\label{phi-correction_sph}
\end{equation}
where
\begin{equation}
    A(t)=\frac{\dot{R}^2\ddot{R}}{(1-\dot{R}^2)^{3/2}}
    \label{eq:A(t)}
\end{equation}
is the amplitude of the leading curvature correction to the field profile.

Examining the above expression we observe that the 
first term corresponds to the correction arising from the boost between the rest frame and the 
laboratory frame. In contrast, the second term is entirely due to the curvature of the worldvolume.

\begin{figure}[hbt!]
    \centering
    \includegraphics[width=0.85\linewidth]{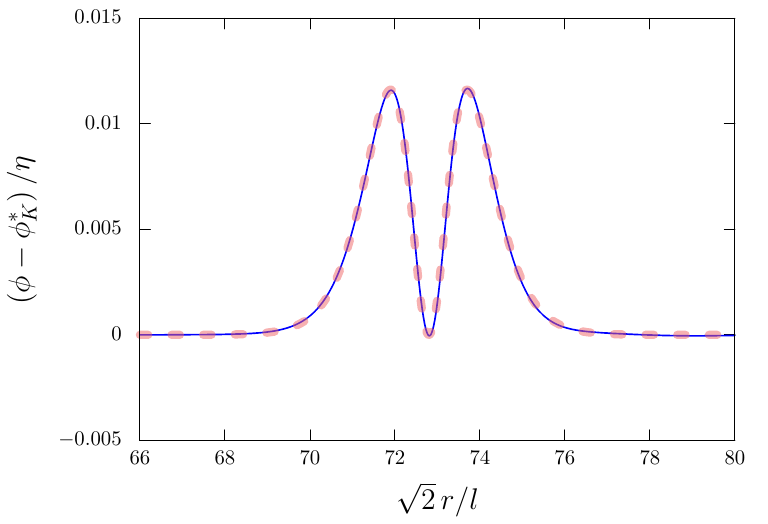}
    \caption{$\phi(r,t)-\phi^*_K(r,t)$ extracted from the field theory simulation at $t\approx39.6\,l$ is shown in blue. The thick pink dashes represent the second term on the right-hand-side of equation (\ref{phi-correction_sph}) evaluated at that time.}
    \label{fig:ears}
\end{figure}

To compare our numerical solution with this analytical prediction, we analyzed the profile during the 
collapse in $3+1$ dimensions. The procedure we follow is to first read the profile from the simulation and subtract from it the first term, i.e. the boosted kink profile, $\phi^*_K(r,t)$, defined as
\begin{equation}
    \phi_K^*(r,t)= \phi_K\qty(\frac{r-R(t)}{\sqrt{1-\dot{R}^2}}).
\end{equation}
The resulting function closely resembles the anticipated correction. See a snapshot of the numerical 
comparison of this correction in \cref{fig:ears}.

It is also interesting to note that this correction increases over time during the collapse and 
becomes quite significant well before the final stages of collapse. This observation agrees with 
the discussions in \cite{Widrow:1989vj} regarding modifications to the equations of motion due to curvature effects.

With the effective action we have developed, we can easily predict the evolution of the amplitude of the curvature correction term as a function of time. Indeed, it is given by \eqref{eq:A(t)}, where the function $R(t)$ can be obtained from the dynamical equations \eqref{rho-equation} and \eqref{EOM} in the 2+1 and 3+1 cases, respectively.  
On the other hand, we can numerically extract the amplitude of such correction as explained in \cref{Numerical-FT}.

\begin{figure}[hbt!]
    \centering
    \includegraphics[width=0.75\linewidth]{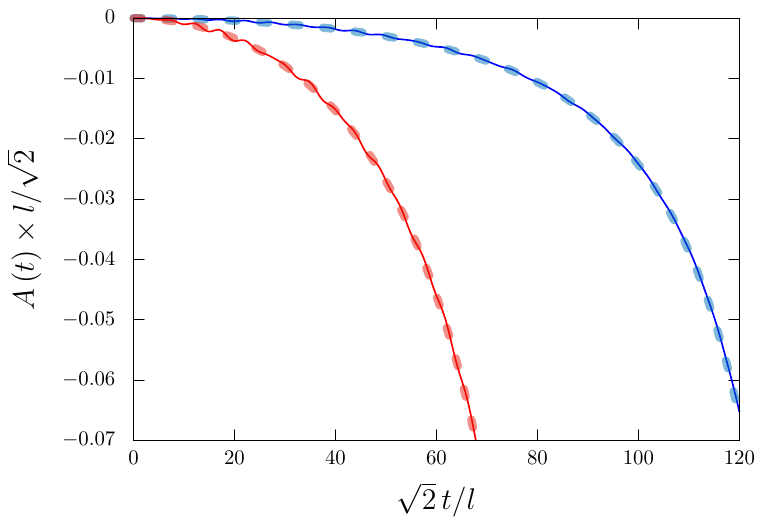}
    \caption{Evolution of the curvature correction amplitude: field theory (solid blue for $2+1$ dimensions and solid red for $3+1$ dimensions) vs effective model (dashed blue for $2+1$ dimensions and dashed red for $3+1$ dimensions).}
    \label{fig:amplitudes_ears}
\end{figure}

We plot the comparison between the predicted evolution and the numerical simulation in  \cref{fig:amplitudes_ears}, with a remarkable agreement between curves.

%%%%%%%%%%%%%%%%%%%%%%%%%%%%%%%%%%%%%%%%%%%%%%%%%
\section{Effective action including Bound States}
\label{sec:boundstate_eff}
%%%%%%%%%%%%%%%%%%%%%%%%%%%%%%%%%%%%%%%%%%%%%%%%%

As shown in the preceding sections, the numerical evolution of the solitons from lattice field theory 
simulations agrees remarkably well with the predictions of our derived effective action. However, this 
agreement deteriorates in the later stages of the wall evolution. A plausible explanation for this deviation 
might be the excitation of new perturbative modes apart from the zero mode at this point. Capturing their 
effects requires deriving the effective action for these modes and their interactions with the zero mode that 
describes the soliton’s position. In this section, we take on that task.

To derive an effective action that includes additional modes, we introduce extra degrees of freedom by expanding the field as $\phi(x)=\phi_K(u)+\chi(u,v)$. Here, as before, $\phi_K(u)$ 
represents the static kink profile as a function of transverse coordinates in its local rest frame, while $\chi(u,v)$ captures the new contributions. For consistency, we impose the constraint\footnote{Here we follow the method outlined in \cite{LinLowe83,Zia:1985gt}.}
\begin{equation}
    \int \phi'_k(u)\chi(v,u) du = 0 \quad~~~~ \forall v~,
\label{Gold_constraint}
\end{equation}
to avoid the double counting of Goldstone degrees of freedom (i.e. the perturbation $\chi$ must not overlap 
with the zero mode). Then, the effective action can be formally written as
\begin{equation}
    S=S^{(0)}_{\rm eff}+\int \chi \frac{
    \delta S
    }{\delta\phi}\eval_{\phi=\phi_K(u)} d^Dx+\frac{1}{2}\int \chi ^2\frac{
    \delta S
    }{\delta^2\phi}\eval_{\phi=\phi_K(u)}d^Dx+\order{\chi^3}~,
    \label{act_expansion}
\end{equation}
where $S^{(0)}_{\rm eff}$ is part that involves just the Goldstone degrees of freedom, given by (\ref{effective_Goldstone}). On the other hand, the tadpole ($\order{\chi})$ terms are generated by the failure of $\phi_K(u)$ to satisfy the full equation of motion. They are given by
\begin{equation}
    \frac{
    \delta S
    }{\delta\phi}\eval_{\phi=\phi_K}=\qty[\pdv{\lag}{\phi}-\partial_\mu\pdv{\lag}{\partial_\mu\phi}]\eval_{\phi=\phi_K}=\pdv{V}{\phi}\eval_{\phi=\phi_K}-\left.\partial_\mu\qty(\eta^{\mu\nu}e_\nu^\alpha\nabla_\alpha\phi)\right|_{\phi=\phi_K}~.
\end{equation}
Now, since 
\begin{equation}  \partial_\mu(\eta^{\mu\nu}e_\nu^\alpha\nabla_\alpha\phi)|_{\phi=\phi_K}= \eta^{\mu\nu}(\partial_\mu e^u_{\nu}\phi_K'+e^u_\mu e^u_\nu \phi''_K)=\eta^{\mu\nu}e_\mu^a\nabla_ae^u_\nu \phi'_K+\phi''_K~,
\end{equation}
and given that $\phi''_K=\partial V/\partial\phi|_{\phi_K}$, and $\nabla_a e^u_\mu=\nabla_an_\mu=-K_a^b\nabla_b\xi_\nu=-K^b_a\delta_{b\nu}$, we have
\begin{equation}
    \frac{
    \delta S
    }{\delta\phi}\eval_{\phi=\phi_K(u)}={\rm Tr}\qty(\frac{\bm{K}}{1-u\bm{K}})~\phi'_K(u),
\end{equation}
where we have used that $e^b_i$ is the inverse matrix of $e^i_a=\delta_ a^i-uK_ a^i$. Therefore, the tadpole terms for $\chi$ in the effective action are gathered in
\begin{equation}
    S^{(1)}_{\rm eff}=-\int d^Dv\sqrt{-\gamma}\int du \det{1-u\bm{K}}~{\rm Tr}\qty(\frac{\bm{K}}{1-u\bm{K}})\phi'_K(u)\chi(u,v)~~.
\end{equation}

Let us now look at the third term in the expansion \eqref{act_expansion}. The second functional derivative is given by
\begin{equation}
    \frac{\delta^2\lag}{\delta\phi^2}=-\eta^{\mu\nu}\partial_\mu\partial_\nu -V''\big|_{\phi_K}~,
\end{equation}
so

\begin{equation}
        S^{(2)}_{\rm eff}=-\frac{1}{2}\int d^Dv\sqrt{-\gamma}\int du \det{1-u\bm{K}}\qty[\chi G^{\alpha\beta}\nabla_{\alpha\beta}\chi +V''\big|_{\phi_K}\chi^2].
        \label{Eff-2}
\end{equation}

%%%%%%%%%%%%%%%%%%%%%%%%%%%%%%%%%%%
\subsubsection{Small u expansion}
%%%%%%%%%%%%%%%%%%%%%%%%%%%%%%%%%%%

Let us assume that we can write the extra term in the field ansatz in the following way, $\chi(u,v)=\theta(v)\sigma(u)$, where $\sigma(u)$ denotes the profile of the mode discussed here while $\theta(v)$ parametrizes the amplitude of the mode on the worldsheet theory. This means that in order to satisfy the constraint in Eq. (\ref{Gold_constraint}) we must impose that 
\begin{equation}
    \int du ~\phi'_K(u)\sigma(u) = 0~,
\end{equation}
In particular, we may choose any anti-symmetric bound state of the static kink (if it exists) and this constraint will be automatically satisfied during its evolution, as the zero mode is always symmetric with respect to the kink center. An example of a mode with this properties in $\lambda \phi^4$
has been already discussed in the introduction, where we introduced the bound state in this model, the {\it shape mode}.

Then, the first non trivial contribution to the small $u$ expansion of the tadpole term is
\begin{equation}
    S^{(1)}_{\rm eff}\approx \kappa \int d^Dv \sqrt{-\gamma}~\mathcal{R}~\theta~,
\end{equation}
where
\begin{equation}
    \kappa =\int du \, u~\phi'_K(u) ~\sigma(u) ~,
    \label{kappa-eq}
\end{equation}
and $\mathcal{R}$ is the Ricci scalar of the induced metric, $\mathcal{R}=K^2-K_{a}^bK_{b}^a$. The small $u$ expansion of the second order term reads
\begin{equation}
     S^{(2)}_{\rm eff}\approx -\frac{1}{2}\int d^Dv\sqrt{-\gamma}\qty[\lambda \gamma^{ab}\nabla_a\theta\nabla_b\theta +m^2\theta^2]~,
\end{equation}
which is the action of a free, massive scalar field on the worldsheet, with
\begin{equation}
    \lambda=\int du ~\sigma^2(u)~~~~\qquad , ~~~~~~ m^2=\int \qty(V''\big|_{\phi_K}\sigma^2+(\sigma')^2)du~.
\end{equation}
Now, if we choose $\sigma$ to be a normal mode of the static kink solution, normalized to unity, then $\lambda=1$. Moreover, its mass $m_*$ satisfies the following identity:
\begin{equation}
    m_*\sigma=-\sigma''+V''\big|_{\phi_K}\sigma~.
\end{equation}
Integrating by parts the second derivative term and substituting back into the definition of $m^2$, the effective action becomes
\begin{equation}
         S^{(2)}_{\rm eff}\approx -\frac{1}{2}\int d^Dv\sqrt{-\gamma}\qty[\gamma^{ab}\nabla_a\theta\nabla_b\theta +m_*^2\theta^2]~.
\end{equation}

Putting everything together, we arrive at

\beq
\label{S_total}
S \approx S^{(0)}_{\rm eff} + \kappa \int d^Dv \sqrt{-\gamma}~\mathcal{R}~ \theta - \frac{1}{2}\int d^Dv\sqrt{-\gamma}\qty[\gamma^{ab}\nabla_a\theta\nabla_b\theta +m_*^2\theta^2].
\eeq
 
This is a remarkable result, demonstrating that the bound state previously identified in
perturbative analyses of excitations around a flat domain wall, denoted here by the amplitude 
$\theta$, couples non-minimally to the Goldstone zero mode via the worldsheet's Ricci scalar. This matches perfectly with the conjecture of such coupling proposed in \cite{Blanco-Pillado:2022rad}.

Finally, it is interesting to consider what this type of effective action reveals about the coupling 
between the soliton worldsheet and the massive radiative modes. Based on the reasoning in this section, 
one can deduce that the effective action for these modes takes a similar form to that in Eq. (\ref{Eff-2}). Following 
analogous steps, one would conclude that these modes couple at the lowest order directly to the Ricci 
scalar of the domain wall worldsheet. This observation offers a promising avenue for studying the 
radiation of massive modes from these solitons. We leave the exploration of these interesting ideas 
for a future publication.

%%%%%%%%%%%%%%%%%%%%%%%%%%%%%%%%%%%%%%%%%%%%%%%%%%%%%%%%%%%%%%%%%%%%%%%%%%%%%%%%%%%%%%
\section{Numerical Evolution of the effective action with a bound state}
\label{Sec:numerical_boundstate}
%%%%%%%%%%%%%%%%%%%%%%%%%%%%%%%%%%%%%%%%%%%%%%%%%%%%%%%%%%%%%%%%%%%%%%%%%%%%%%%%%%%%%%

In this section, we explore two significant consequences of the bound state's inclusion in 
our extended effective action. First, we analyze the parametric instability in the wall's 
position arising from the non-zero amplitude of the bound state. Second, we investigate the 
bound state's amplification during a domain wall collapse and its backreaction on the position of
the wall. In both cases, we confirm the validity of this effective action by demonstrating 
its agreement with the full numerical field theory configurations.

%%%%%%%%%%%%%%%%%%%%%%%%%%%%%%%%%%%
\subsection{Parametric Instability}
%%%%%%%%%%%%%%%%%%%%%%%%%%%%%%%%%%%

\begin{figure}[hbt!]
    \centering
    \includegraphics[width=0.7\linewidth]{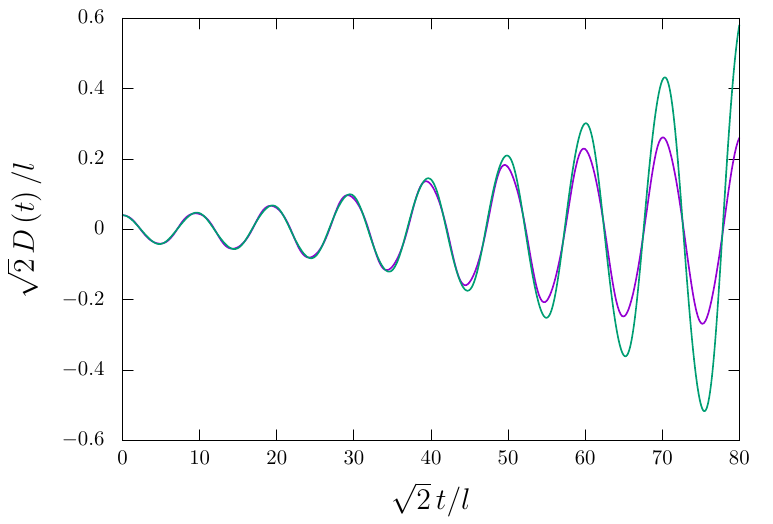}
    \caption{Parametric resonance of the zero mode in the $\lambda\phi^{4}$ domain wall string. The purple curve represents the field theory simulation, and the green curve the prediction from the Mathieu equation obtained in the effective model \eqref{S_total}, that is, equation (\ref{eq:amplitude zero mode}).}
    \label{fig:resonance}
\end{figure}

For the $\lambda\phi^{4}$ domain wall string, the effective action \eqref{S_total} predicts that 
the amplitude $D\left(t\right)$ of small transverse displacements in the form of standing waves of
frequency $\omega_{0}$ obeys a Mathieu equation of the form
\begin{equation}
\ddot{D}\left(t\right)+\omega_{0}^{2}\left[1-\frac{2\kappa m_{*}}{\mu}\theta\left(t\right)\right]D\left(t\right)=0 ~,
\label{eq:amplitude zero mode}
\end{equation}
to lowest order in both $D$ and $\theta$. Here, $\mu$ is the energy per unit length of the string, $m_*$ is the
mass of the bound state and $\kappa$ is the coupling constant computed following Eq. (\ref{kappa-eq}). In \cref{fig:resonance} we compare the solution of this equation to the amplitude of the zero mode directly read from a field theory simulation where the parametric resonance took place\footnote{In order for the resonance to occur rapidly in the simulation, the string was initialized with the shape of a standing wave with the resonant frequency $\omega_{0}\approx m_{*}/2$ and a tiny amplitude. More details, including the numerical computation of the amplitude of the zero mode, can be found in \cite{Blanco-Pillado:2022rad}.}. It is clear that the first oscillations in field theory are successfully predicted by the coupling between $\theta$ and the Ricci scalar of the string worldsheet provided by the effective theory at the lowest order. 

This analysis confirms the presence of such coupling in our theory but more importantly it also suggests such couplings are likely a generic feature of models with bound states. It also establishes that the parametric instability identified in the $\lambda\phi^{4}$ model in 
\cite{Blanco-Pillado:2022rad} is not an isolated phenomenon but rather a universal characteristic 
within this class of models.

%%%%%%%%%%%%%%%%%%%%%%%%%%%%%%%%%%%%%%%%%%%%%%%%%%%%%%%%%%%%%%%%%%%%%%%%%
\subsection{Effects of the bound state amplitude in the collapsing wall}
%%%%%%%%%%%%%%%%%%%%%%%%%%%%%%%%%%%%%%%%%%%%%%%%%%%%%%%%%%%%%%%%%%%%%%%%%

As argued in sec. \ref{sec:numerics_effective}, the radius of the domain wall follows a trajectory which coincides with the Nambu-Goto prediction up to a small correction. We have also shown that the first nontrivial correction (for the $\phi^4$ model in 3+1 dimensions, for instance) comes from the term proportional to the Ricci curvature of the worldsheet in the effective Lagrangian \eqref{3+1-Eff}. However, at certain point, the effect of additional bound modes becomes important and the effective action must be completed with the additional terms in \eqref{S_total}. 

\begin{figure}[hbt!]
    \centering
    \includegraphics[scale=0.95]{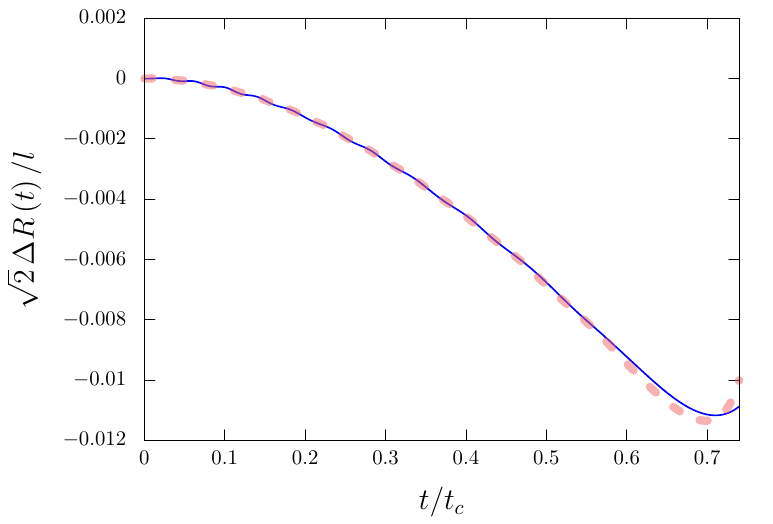}
    \caption{$R_{NG}(t)-R(t)$ for the $\lambda\phi^{4}$ domain wall in $3+1$ dimensions (compare with \cref{fig:spherical-wall-1}). The solid blue line corresponds to the numerical field theory result, and the dashed pink line is the prediction of the effective model (\ref{S_total}). Time is displayed in units of the collapse time, which is approximately given by $t_{c}\approx92.7\,l$ for an initial radius $R_{0}=\left(100/\sqrt{2}\right)l$.}
    \label{fig:mass}
\end{figure}

In \cref{fig:mass} we see that the effect of the bound mode is a small 'bounce' of the $R_{NG}-R$ curve close to the collapse time. When comparing with the numerical simulation, we see that by simply considering the $\phi^4$ static kink shape mode, the dynamics is well reproduced. In particular, the position of the 'bounce' is accurately predicted to take place at around $0.7$ times $t_c$.

The amplitude of the bound state, as predicted by our effective field theory, grows significantly as the wall 
approaches collapse. This behavior can be intuitively understood by picturing the bound state's amplitude 
as a scalar field living on the collapsing wall’s worldsheet. Drawing an analogy with a scalar field in a 
collapsing universe, the contraction naturally leads to an increase in the amplitude of an oscillating scalar 
field. This growth also signals a significant energy transfer to the bound state, which in turn backreacts 
on the wall’s motion, explaining the subtle ``bounce'' effect observed in our simulations.

Having established the impact of collapse on the bound mode’s amplitude, it is reasonable to suspect that 
other continuum modes might also be excited during this process. These excitations could influence the 
wall’s dynamics further, potentially signaling the limits of the effective theory’s applicability.

In fact, the final stages of collapse clearly extend beyond the realm of effective field theory. This process 
can be viewed as a kink-antikink collision and will undoubtedly excite radiative modes not captured by our 
current framework. Describing this final process requires the introduction of non-local terms in the action
to adequately account for the interactions between various regions of the domain wall's worldvolume. Understanding 
this phenomenon and its relation to our effective field theory is an interesting problem, 
albeit one that lies outside the scope of the present paper.\\

%%%%%%%%%%%%%%%%%%%%%%%%%%%%%%%%%%%%%%%%%%%
\section{Conclusions}
\label{Sec:Conclusions}
%%%%%%%%%%%%%%%%%%%%%%%%%%%%%%%%%%%%%%%%%%%

In this work we have presented a systematic method for obtaining the most general effective action for codimension-1 defects (domain walls or $D-1$ branes) from the underlying UV field theory, in terms of the curvature invariants associated to its worldvolume metric and their couplings to an arbitrary number of scalar fields living on the worldvolume. In particular, we show that the effective action for the lowest energy degree of freedom, namely, Goldstone fluctuations along the dimension transverse to the wall, can be written in terms of the so-called DBI-Galileon terms \cite{deRham:2010eu}. Note that this requirement highly constrains the effective action, since the number of different non-trivial DBI-Galileon terms is finite for any given number of spacetime dimensions.
Therefore, our method is consistent both with the Goldstone EFT philosophy and with the general requirement that the Hamiltonian dynamics of the Goldstone fields is well defined, i.e. the absence of ghost-like degrees of freedom in the classical limit.

The method we describe is based on the definition of a set of normal coordinates adapted to the worldvolume of the brane, which allowed us to express several curvature corrections in the form of a coordinate change to the 'kink frame', in terms of which the functional form of the solution is that of the static wall. This method not only has allowed us to obtain the effective action, but also the corrections to the kink profile in the lab frame, by inverting such coordinate change. Since the kink frame coordinates are only defined in a neighborhood of the wall's worldvolume, inverting the coordinate change can only be accomplished perturbatively in powers of the distance to the wall. The ability to perform such coordinate change in fact defines the range of validity of our effective action. Indeed, we expect the effective action to fail precisely for small curvature radii of the worldvolume, i.e. of order of the width of the wall. This includes not only the regions of large spatial curvature of the wall, but also high velocities. 

We have rigorously tested the predictions for domain wall dynamics in 2+1 and 3+1 dimensions derived from the effective action against numerical results from field theory. The agreement is striking, even when deviations from Nambu-Goto behavior—arising from additional curvature invariants such as the trace of the extrinsic curvature and the Ricci scalar—are small relative to the leading contributions. These comparisons also allowed us to explicitly demonstrate how curvature effects modify the domain wall profile, further validating our methodology.

Going one step further, our approach has also enabled the derivation of the effective action for a bound state living on the worldsheet, including its coupling to the Goldstone mode. We find that the emergence of a non-minimal coupling, involving a term linear in the scalar field and the Ricci scalar, is universal in such systems. A key implication of this coupling is the presence of a parametric resonance, whereby the wall's position can develop oscillatory instabilities solely due to the presence of a uniform bound state. We have also investigated this process numerically and found 
a remarkable agreement between the full field theory dynamics and the one predicted by our effective
action. This clearly validates not only the accuracy of the terms present in the effective theory but also the method for computing its coefficients.

The results presented in this work may play a significant role in cosmological simulations of field 
theory solitons, particularly domain walls. By quantifying the magnitude of the effective action 
coefficients, we can estimate the regimes where these corrections become non-negligible and introduce 
some compensating terms to the equations of motion to deal with them in the simulations. Moreover, the 
coupling to the bound state introduces new intriguing possibilities, potentially manifesting as 
observable effects during numerical evolution. 

It is worth noting, however, that these bound states are only stable at linear order. For long-term dynamics, such as in cosmological settings where solitons may persist over extended periods, it is essential to account for the decay of these bound states. This is particularly relevant for understanding their potential observational consequences in such environments.

Finally, our method also suggests a novel framework for studying the coupling between domain walls and radiation modes. This could shed light on longstanding debates about dominant decay mechanisms for topological defects in cosmology. 

Looking ahead, our approach invites exploration of its applicability to other types of defects. Particularly interesting is its extension to higher co-dimension objects like cosmic strings or its relevance in higher-dimensional theories, such as braneworld scenarios. Indeed, in the case of higher co-dimension objects, the relevant multi-Galileon terms were already found in \cite{Hinterbichler:2010xn} following a bottom-up approach. It would be interesting to re-derive an effective action containing such terms from a higher dimensional field theory that presents higher co-dimension solitons.
Furthermore, it would also be interesting to study the effects of our findings when we couple these objects to gravity, in which case the effective action should contain additional gravity-induced terms \cite{Goon:2011xf,PhysRevLett.106.231102}, that are expected to modify significantly their dynamics when considering curved spacetime backgrounds such a black hole \cite{Christensen:1998hg,Frolov:1998td}.

Another interesting avenue would be to investigate the importance of the derived couplings between the zero modes, bound states and radiation in the quantum description of solitons. Indeed, the interplay between zero modes and radiation has been recently shown to be important for quantum solitons in 1+1 dimensions \cite{Evslin:2021nsi,Evslin:2022wyx,Evslin:2022xmp} and similarly promising results are being developed for higher dimensional solitons \cite{Evslin:2024sup,Ogundipe:2024ibv,Berezhiani:2024pub}. While we have not addressed these questions in detail here, ongoing work is already investigating these exciting directions.

\begin{acknowledgments}
We are grateful to Igor Bandos, Ken D. Olum and Oriol Pujolas for stimulating discussion.  This work has been supported in part by the PID2021-123703NB-C21 grant funded by MCIN/AEI/10.13039/501100011033/and by ERDF;“ A way of making Europe”; the Basque Government grant (IT-1628-22) and the Basque Foundation for Science (IKERBASQUE). JQ has been supported in part by Spanish Ministerio
de Ciencia e Innovaci\'on (MCIN) with funding from the
European Union NextGenerationEU (PRTRC17.I1) and
the Consejer\'ia de Educaci\'on, Junta de Castilla y Le\'on,
through QCAYLE project, as well as the grant
PID2023-148409NB-I00 MTM, and the project Programa C2 from the University of Salamanca.
DJA is supported in part by National Science Foundation grants PHY-2110466 and PHY-2419848.
\end{acknowledgments}

\appendix
%%%%%%%%%%%%%%%%%%%%%%%%%%%%%%%%%%%%%%%%%%%
\section{Adapted coordinate system}
\label{u-expansion}
%%%%%%%%%%%%%%%%%%%%%%%%%%%%%%%%%%%%%%%%%%%
In the main part of the text we introduced a new coordinate system for flat space
adapted to the domain wall worldvolume. Here we want to study the relation between this new system
and the usual one in more detail. 

Let us first consider a wall of arbitrary shape parametrized by the function $z = \psi(y^a)$ where the flat spacetime original coordinates are given by $x^{\mu} = (y^a, z)$. Now let us define a new coordinate system $\zeta^{\mu}= (v^a,u)$ such that

\begin{equation}
   x^\mu=\xi^\mu(v)+un^\mu(v)~,
\end{equation}
where $\xi^a=v^a$, $\xi^{d+1}=\psi(v)$, and $n^\mu$ is the $D$-dimensional unit vector normal to the wall at each point in $v$.

In this appendix, we are interested in computing the inverse coordinate transformation, $\zeta^\mu\equiv \zeta^\mu(y^a,z)$, and, in particular, the surfaces $u(z,\psi(y))$, as an expansion in powers of $(z-\psi(y)).$ Explicitly, the coordinate change is
\beqa
y^a(v^a,u)&=&\xi^a+ u n^a=v^a-\frac{u}{\sqrt{-\gamma}}\nabla^a \psi \label{change1}~,\\
z(v^a,u)&=&\xi^D+u n^D=\psi+\frac{u}{\sqrt{-\gamma}}~.
\label{change2}
\eeqa
Note that in the rhs of both (\ref{change1}) and (\ref{change2}), all functions are evaluated only on $v^a$, and hence the dependence on $u$ is explicit.

Furthermore, since the newly defined coordinate system will always be well defined on a sufficiently small neighborhood of the $u=0$ surface, we will expand
\begin{equation}
    z-\psi(y)=\alpha(y) u+\beta (y) u^2+\cdots
    \label{z_exp}
\end{equation}
where there is no dependence on $v$, as $v$ can be found as a funtion of $u$ and $y^a$ at any given order in powers of $u$ via (\ref{change1}). Hence, we can truncate the expansion \eqref{z_exp} to the desired order in $u$, and invert it to find
\begin{equation}
    u(y,z)=\frac{1}{\alpha(y)}(z-\psi(y))-\frac{\beta(y)}{\alpha(y)^3}(z-\psi(y))^2+\cdots
    \label{inv_exp}
\end{equation}
Now the task is reduced to computing the coefficients $\alpha(y)$ and $\beta(y)$ in the above expansion.

%%%%%%%%%%%%%%%%%%%%%%%%%%%%%%%%%%%%%%%%%%%
\subsubsection{First order}
%%%%%%%%%%%%%%%%%%%%%%%%%%%%%%%%%%%%%%%%%%%

Let us start by expanding the function $\psi(v)$ around $v^a=y^a$ (i.e. $u=0$)\footnote{We remind that indices in this appendix are always contracted using the flat spacetime metric $\eta_{ab}$.}:
\beqa
    \psi(v)&=&\psi(y)+\nabla_ a\psi(v)\eval_{y}\!\!(v-y)^a+\frac{1}{2}\nabla_{ab}\psi(v)\eval_y(v-y)^a(v-y)^b+\cdots\notag\\
    &=&\psi(y)+\partial_a\psi(y)(v-y)^a+\frac{1}{2}\partial_{ab}\psi(y)(v-y)^a(v-y)^b+\cdots
\eeqa
On the other hand, from (\ref{change1}), 
\begin{equation}
    (v-y)_a=\frac{u}{\sqrt{-\gamma(v)}}\nabla_ a\psi(v)=\frac{u}{\sqrt{-\gamma(y)}}\partial_ a\psi(y)+\order{u^2}.
    \label{vu_o1}
\end{equation}
Therefore, from (\ref{change2}),
\begin{equation}
    \frac{u}{\sqrt{-\gamma(v)}}=z-\psi(v)=z-\psi(y)-\partial_a\psi(y)(v-y)^a=z-\psi(y)-\frac{u}{\sqrt{-\gamma(y)}}\partial_a\psi(y)\partial^a\psi(y)
\end{equation}
from where we deduce
\begin{equation}
    z-\psi(y)=u\frac{1+\partial_a\psi(y)\partial^ a\psi(y)}{\sqrt{-\gamma(y)}}+\order{u^2}\equiv \sqrt{-\gamma(y)} u+\order{u^2}
\end{equation}
and inverting this relation we find
\begin{equation}
    u=\frac{z-\psi(y)}{\sqrt{-\gamma(y)}}+\order{(z-\psi)^2}
\end{equation}
from where we can directly read the first coefficient
\begin{equation}
    \alpha(y)=\sqrt{-\gamma(y)}
\end{equation}
as we already expected from the discussion in the previous sections
.
%%%%%%%%%%%%%%%%%%%%%%%%%%%%%%%%%%%%%%%%%%%
\subsubsection{Second order }
%%%%%%%%%%%%%%%%%%%%%%%%%%%%%%%%%%%%%%%%%%%

First of all, we need the derivative of $\psi$ at next order. To ease the notation, we will omit the argument of functions of the flat coordinates $y^a$, but will keep the argument of functions of the adapted coordinates $v^a$. Then, we have
\begin{align}
    \nabla_ b\psi(v)&=e^a_b\partial_a\psi(v)=\qty[\delta_b^a-u\nabla^a\qty(\frac{\nabla_b \psi(v)}{\sqrt{-\gamma(v)}})]\partial_a \psi(v)=\notag\\
    &=\qty[\eta_{ab}-\frac{u}{\sqrt{-\gamma}}\partial_{ab}\psi +\frac{1}{2}\frac{u}{|\gamma|^{3/2}}\partial_b\psi\partial_a\gamma+\order{u^2}]\partial^a[\psi+\partial_c\psi(v-y)^c+\cdots]=\notag\\
    &=\partial_b\psi-\frac{u}{\sqrt{-\gamma}}\partial_{ab}\psi\partial^a\psi +\frac{1}{2}\frac{u}{|\gamma|^{3/2}}\partial_b\psi\partial_a\gamma\partial^a\psi+ u\partial_b\qty(\frac{\partial_a\psi\partial^a\psi}{\sqrt{-\gamma}})+\order{u^2}=\notag\\
    &= \partial_b\psi+\frac{u}{\sqrt{-\gamma}}\partial_{ab}\psi\partial^a\psi +\order{u^2}.
\end{align}
For the metric determinant, we have
\begin{equation}
    \frac{1}{\sqrt{-\gamma(v)}}=\frac{1}{\sqrt{-\gamma}}+\partial_a\qty(\frac{1}{\sqrt{-\gamma}})(v-y)^a+\cdots = \frac{1}{\sqrt{-\gamma}}-\frac{1}{2}\frac{u}{|\gamma|^2}\partial_ a \psi \partial^ a |\gamma|+\order{u^2}.
\end{equation}
Thus, we can write the next order in \eqref{vu_o1}:
\begin{align}
    (v-y)^a&=\frac{u}{\sqrt{-\gamma(v)}}\nabla_ a\psi(v)=\notag\\
    &=u\qty(\frac{1}{\sqrt{-\gamma}}-\frac{1}{2}\frac{u}{|\gamma|^2}\partial_ a \psi \partial^ a |\gamma|+\order{u^2})\qty(\partial_a\psi+\frac{u}{\sqrt{-\gamma}}\partial_{ab}\psi\partial^b\psi +\order{u^2})=\notag\\
    &= \frac{u}{\sqrt{-\gamma}}\partial_ a\psi+u^2\qty[\frac{\partial_{ab}\psi\partial^b\psi}{|\gamma|}-\frac{1}{2}\frac{\partial_ a \psi \partial_ b \psi \partial^ b |\gamma|}{|\gamma|^2}]+\order{u^3}.
\end{align}
Therefore, we can write \eqref{change2} up to second order in $u$ as:
\begin{align}
    z&=\psi(v)+\frac{u}{\sqrt{-\gamma(v)}}=\notag\\
    &=\psi +\partial_ a\psi (v-y)^a+\frac{1}{2}\partial_{ab}\psi(v-y)^a(v-y)^b+u\qty(\frac{1}{\sqrt{-\gamma}}-\frac{1}{2}\frac{u}{g^2}\partial^ a \psi \partial_ a |\gamma|)+\order{u^3}=\notag\\
    &=\psi +u\frac{(1+\partial^a\psi\partial_a\psi)}{\sqrt{-\gamma}}+u^2\qty{\frac{\partial^ a\psi\partial_{ab}\psi\partial^b\psi}{|\gamma|}-\frac{1}{2}\frac{(1+\partial^ a \psi \partial_a\psi)\partial^ b \psi \partial_ b |\gamma|}{|\gamma|^2}+\frac{1}{2}\frac{\partial^ a\psi\partial_{ab}\psi\partial^b\psi}{|\gamma|}}+\order{u^3}=\notag\\
    &=\psi+\sqrt{-\gamma}u+\frac{1}{2}\frac{\partial^ a\psi\partial^b\psi\partial_{ab}\psi}{|\gamma|}u^2+\order{u^3},
\end{align}
from where we can read the second coefficient:
\begin{equation}
    \beta(y)=\frac{1}{2}\frac{\partial^a\psi\partial^b\psi\partial_{ab}\psi}{|\gamma|},
\end{equation}
hence we can write the inverse coordinate transformation \eqref{inv_exp} as
\begin{equation}
    u=\frac{z-\psi}{\sqrt{1+\partial^a\psi\partial_a\psi}}-\frac{1}{2}\frac{(z-\psi)^2}{(1+\partial^a\psi\partial_a\psi)^{5/2}}\partial^a\psi\partial^b\psi\partial_{ab}\psi+\order{(z-\psi)^3}.
\label{Expansion_u}
\end{equation}
%Note the sign difference with respect to \cite{LinLowe83} in the definition of $f$  (here $\psi$). We use the convention of \cite{Zia:1985gt}, where there is a sign error.  

%%%%%%%%%%%%%%%%%%%%%%%%%%%%%%%%%%%%%%%%%%%%%%%%%
\section{Numerical Field Theory Implementation}
\label{Numerical-FT}
%%%%%%%%%%%%%%%%%%%%%%%%%%%%%%%%%%%%%%%%%%%%%%%%%

In this appendix, we will provide the details of the lattice field theory simulations carried out in this 
work. In both the $2+1$ and the $3+1$ dimensional settings, the symmetry of the problem (cylindrical or 
spherical, respectively) allowed us to solve the field theory equation of motion in a $1+1$ dimensional
lattice. In order to probe small deviations from the Nambu-Goto action, we used a huge number of points 
($\sim 10^{5}$) in our lattice.

\subsection{Equations of motion and boundary conditions}
In order to numerically solve the equation of motion, we first define the following dimensionless variables: $\tilde{\phi}=\phi/\eta$, $\tilde{x}^{\mu}=\sqrt{\lambda}\eta x^{\mu}$. In terms of these new variables, the equation of motion coming from the field theory action (\ref{FT-action}) is free of parameters. Likewise, the time step $\Delta t$, lattice spacing $\Delta x$ and length of the simulation box $L$ are also made dimensionless upon multiplication by $\sqrt{\lambda}\eta$. In the following, all the variables appearing in the text will be dimensionless, but we will drop the tildes for the sake of simplicity in the notation.\\

In both the $2+1$ and $3+1$ cases, we chose $\Delta t=10^{-4}$, $\Delta x=4\times 10^{-4}$ and $L=140$, and the wall starts from rest with initial radius $R_{0}=100$. Moreover, we parallelized our code in order to handle the huge number of lattice points.\\

The dimensionless equations of motion for the $\lambda\phi^{4}$ theory read\\
\begin{equation}
\frac{\partial^{2}\phi}{\partial t^{2}}-\frac{\partial^{2}\phi}{\partial r^{2}}-\frac{1}{r}\frac{\partial\phi}{\partial r}+\phi^{3}-\phi=0~,
\label{eq:2+1 eq}
\end{equation}
and 
\begin{equation}
\frac{\partial^{2}\phi}{\partial t^{2}}-\frac{\partial^{2}\phi}{\partial r^{2}}-\frac{2}{r}\frac{\partial\phi}{\partial r}+\phi^{3}-\phi=0~,
\label{eq:3+1 eq}
\end{equation}
in $2+1$ and $3+1$ dimensions, respectively. In both cases, we solved the equation using the staggered leapfrog method and nearest-neighbors discretization for the spatial derivatives, with the following initial conditions:
\begin{equation}
\phi\left(t,r\right)=\tanh\left(\frac{r-R_{0}}{\sqrt{2}}\right)\,\,\,~~~~;~~~~\,\,\,\dot{\phi}\left(t,r\right)=0\,.
\label{eq:initial condition}
\end{equation}
As we mentioned in the main part of the text, we take this initial condition for large values
of $R_0$ compared to the soliton's thickness in an attempt to start with a configuration that
is close to an exact solution of the non-linear equations of motion.

Regarding the boundary conditions, we imposed $\partial_{r}\phi\left(t,r=0\right)=0$ and an absorbing condition at $r=L$:
\begin{equation}
\frac{\partial\phi}{\partial t}+\frac{\partial\phi}{\partial r}\,\eval_{r=L}=0\,.
\label{eq:absorbing-bc}
\end{equation}
\\
Similarly, for the $2+1$ dimensional $\lambda\phi^{6}$ simulations, we make the spacetime coordinates dimensionless by the rescaling $\tilde{x}^{\mu}=x^{\mu}/\ell$, with $\ell=\sqrt{2/\lambda}\,\eta^{-2}$. In this case, the dimensionless equation of motion reads
\begin{equation}
\frac{\partial^{2}\phi}{\partial t^{2}}-\frac{\partial^{2}\phi}{\partial r^{2}}-\frac{1}{r}\frac{\partial\phi}{\partial r}+\left(\phi^{3}-\phi\right)\left(3\phi^{2}-1\right)=0\,.
\label{eq:2+1 eq lambdaphi6}
\end{equation}

\subsection{Amplitude of the curvature perturbation and the bound state}
Here we will explain how to read the amplitude of the curvature perturbation and the bound state from the field theory simulations.\\\\
The former corresponds to the amplitude $A\left(t\right)$ appearing in equation (\ref{phi-correction_sph}). In order to read it directly in the simulation, we subtract the boosted profile from both sides of the equation and integrate over the radial direction $r$ to get
\begin{equation}
A\left(t\right)\approx\frac{\int_{0}^{L}dr\left[\phi\left(t,r\right)-\phi_{K}\left(\frac{r-R(t)}{\sqrt{1-\dot{R}^{2}(t)}}\right)\right]}{\int_{0}^{L}dr\left[\frac{r-R(t)}{\sqrt{1-\dot{R}^2(t)}}\right]^2\phi_K'\qty(\frac{r-R(t)}{\sqrt{1-\dot{R}^2(t)}})}\,.
\label{eq:amplitude curvature perturbation projection}
\end{equation}
Note that this calculation requires not only the position of the wall but also its velocity. The huge number of lattice points in our simulations allowed us to extract the position of the wall with high precision by identifying the two points where the field changed sign and finding its zero by linear interpolation. At every time step, we stored this position, $R(t)$, as well as the position obtained at the previous time step, $R(t-\Delta t)$. Then, the velocity is simply obtained as $\dot{R}(t)=\left[R(t)-R(t-\Delta t)\right]/\Delta t$.\\\\
The computation of the amplitude of the bound state, $\theta(t)$, is similar. Assuming that the field configuration is approximately given by
\begin{equation}
\phi\left(t,r\right)\approx\phi_{K}\left(u\right)+\theta(t)\sigma(u)+f(u)\,,
\label{eq:bound state projection 1}
\end{equation}
where 
\begin{equation}
u(t,r)=\frac{1}{\sqrt{2}}\frac{r-R(t)}{\sqrt{1-\dot{R}^{2}(t)}}\left[1+A(t)\frac{r-R(t)}{\sqrt{1-\dot{R}^{2}(t)}}\right]\,,
\label{eq:u bound state projection}
\end{equation}
and $\sigma(u)$ and $f(u)$ denote, respectively, the bound state and other modes orthogonal to it (see for example \cite{Vachaspati:2006zz} to obtain the function form of these profile functions), one can find $\theta(t)$ as
\begin{equation}
\theta(t)\approx\frac{\int dr\left[\phi(t,r)-\phi_{K}(u)\right]\sigma(u)}{\int dr\sigma^{2}(u)}\,.
\label{eq:bound state projection 2}
\end{equation}

Here, one has to be careful about the limits of integration. Indeed, $\phi_{K}(u)$ and $\sigma(u)$ vanish for two different values of $r$, and way before the collapse time, these two roots are positive. In other words, our coordinate system centered at the wall stops working relatively soon, so one has to calculate the integrals over a sufficiently small range of $r$ near the wall.

%%%%%%%%%%%%%%%%%%%%%%%%%%%%%%%%%%%%%%%%%%%%%%%%%
\section{Details of the effective dynamics}
\label{app:details}
%%%%%%%%%%%%%%%%%%%%%%%%%%%%%%%%%%%%%%%%%%%%%%%%%
We consider a spherical domain wall, whose center describes a worldvolume embedded on flat spacetime through
\begin{equation}
    x^\mu(t,\theta,\phi)=(t,R(t),\theta,\phi)
\end{equation}
in spherical coordinates, in which the spacetime metric can be written
\begin{equation}
    ds^2=-dt^2+dr^2+r^2(d\theta^2+\sin^2\theta d\phi^2).
\end{equation}
The metric induced on the worldvolume is given by 
\begin{equation}
    \gamma_{ab}=\eta_{\mu\nu}e^\mu_a e^\nu_b=\mqty(
 \dot R(t)^2-1 & 0 & 0 \\
 0 & R(t)^2 & 0 \\
 0 & 0 & \sin ^2\theta R(t)^2 )
\end{equation}
where $e^\mu_a=\partial_a x^\mu$, and the normalized four-vector orthogonal to such hypersurface is 
\begin{equation}
    n_\mu=\qty(-\frac{\dot R(t)}{\sqrt{1-\dot{R}(t)^2}},\frac{1}{\sqrt{1-\dot R(t)^2}},0,0),
\end{equation}
i.e. $n_\mu e^\mu_a=0$, and $n_\mu n^\mu=1$.

With this ansatz, we can compute the trace of the extrinsic curvature and the Ricci scalar in terms of the bubble radius $R(t)$, so that the effective action becomes:
\begin{equation}
    S_{DW}^{3+1}=-\mu\int  R^2 \sqrt{\dot R^2-1} \qty[1+C_1\frac{ \left(R \ddot R-2 \dot R^2+2\right)}{R \left(1-\dot{R}^2\right)^{3/2}}+C_2\frac{\left(4 R \ddot R-2 \dot R^2+2\right)}{R^2 \left(\dot R^2-1\right)^2}] dt\equiv \int L(R,\dot R,\ddot R) dt
    \label{effective_lag}
\end{equation}
Moreover, the corresponding EOMs must take into account the appearance of higher (second) order derivative terms in the action by including an additional term:
\begin{equation}
    \pdv{L}{R}-\dv{}{t}\pdv{L}{\dot R}+\dv{^2}{t^2}\pdv{L}{\ddot R}=0,
\end{equation}
which yields \cref{EOM} in the main text.
The effective Lagrangian defined in \cref{effective_lag} presents higher derivatives, but it is non-degenerate. Therefore, as expected, the resulting equation of motion does not present higher order derivatives, and the initial value problem is well defined. Also, we note that, in the limit of vanishing coupling constants $c_{1,2}\to 0$, we recover the well known equation for a collapsing spherical domain wall in the Nambu-Goto approximation \cite{Vachaspati:2006zz}:
\begin{equation}
    \ddot R=-2\left(\frac{1-\dot R^2}{R}\right)~.
\end{equation}
This equation presents an analytic solution for a spherical domain wall starting at rest with a radius of $R_0$,
\beq
R(t) = R_0 ~\text{cd}\left(\frac{t}{R_0}, -1\right)
\label{NG 3plus1}
\eeq
in terms of the Jacobi  cd-elliptic function  $\text{cd}(u,m)$ (see chap. 22 in \cite{NIST:DLMF}).

The Nambu-Goto approximation will be also very good for very large initial radii. We expect to find nontrivial curvature corrections only for values of initial radius of order $R_0\sim 10 l$.
Equation \eqref{EOM} cannot be solved analytically, but can easily be integrated numerically. 
We note that, while $C_2\neq 0$ in general, $C_1$ will vanish identically for any $\mathbb{Z}_2$ symmetric potential, such as the $\phi^4$ model.
In such cases, the extrinsic curvature will not appear in the effective action. However, in the most general case, the curvature of the potential around the different vacua will not be the same, and this term should be taken into account. 
In fact, for configurations with $C_1<0$, there exists an equilibrium solution of the effective EOMs \eqref{EOM} satisfying $\Ddot{R}=\dot{R}=0$ at $R_{\rm eq}=-C_1$. Furthermore, by looking at the dynamics of linear perturbations around such point, $R(t)=-C_1+\delta(t)$, we find
\begin{equation}
    \ddot{\delta}=\frac{2 }{3 (C_1^2-2 C_2)}\delta+\order{\delta^2}
\end{equation}
which implies that the condition for a stable equilibrium point is
\begin{equation}
    C_1^2-2C_2<0
\end{equation}
which is \emph{always} true by virtue of the Cauchy-Schwarz inequality. 
In other words, the effective theory predicts the existence of an equilibrium configuration for closed asymmetric domain walls in 3+1 dimensions. However, such a configuration would violate Derrick's scaling theorem \cite{Derrick:1964ww}, hence we expect that the effective action will no longer become appropriate for walls of radius $R\approx \abs{C_1}\sim l$.
An equivalent way of understanding this is by analyzing the potential energy of the wall as a function of its radius. In order to do so, we must rewrite the effective Lagrangian  in the canonical form of kinetic minus potential energy. We will always be able to do so, because the effective Lagrangian $L(\ddot{R},\dot{R},R)$ is non degenerate. Therefore, it must be of the form
\begin{equation}
    L(\ddot{R},\dot{R},R)=F(\dot{R},R)\ddot{R}+G(\dot{R},R)\,.
\end{equation}
Integrating by parts, we can write an equivalent Lagrangian which does not depend on second derivatives:
\begin{equation}
    \tilde L(\dot{R},R)=-\dot{R}\partial_R\int_0^{\dot{R}} F(s,R)ds+G(\dot{R},R)\,.
\end{equation}
Then, the function $V(R)\equiv \tilde{L}(0,R)$ can be interpreted as a potential energy. For the $3+1$ spherical domain wall, we have
\begin{equation}
    V(R)=\mu(2C_2+2C_1 R+ R^2)
\end{equation}
which presents a minimum at $R_{\rm eq}=-C_1$.

In the 2+1 case, with a circular domain wall string parametrized as in \cref{coords2+1} in the main text, only the extrinsic curvature term contributes to the effective action:
\begin{equation}
    S_{DW}^{2+1}=-\mu\int  \rho\sqrt{\dot \rho^2-1} \qty[1+C_1\frac{ \left(1+ \rho\ddot \rho- \dot \rho^2\right)}{\rho \left(1-\dot{\rho}^2\right)^{3/2}}] dt
    \label{effective_2+1}
\end{equation}

The procedure to obtain the EOMs is the same, which straightforwardly yields \cref{rho-equation}.

\bibliography{Bibliography}
\end{document}